\magnification=\magstep1
\settabs 18 \columns
 
\input epsf

\hsize=16truecm

\def\b{\bigskip}
\def\bb{\bigskip\bigskip}

\def\no{\noindent}
\def\r{\rightline}
\def\ce{\centerline}
\def\ve{\vfill\eject}

\def\r{\rightline}

 \def\L{{\cal L}}

\def\harr#1#2{\smash{\mathop{\hbox to .25 in{\rightarrowfill}}
 \limits^{\scriptstyle#1}_{\scriptstyle#2}}}

\def\R{{\cal R}}
\def\V{{\cal V}}

\def\today{\ifcase\month\or January\or February\or March\or April\or
May\or June\or July\or
August\or September\or October\or November\or  December\fi
\space\number\day, \number\year }

\r {HeatandGravIIIApril2010}

\r \today
\bb\bb\bb

\def\Rrm{\hbox{\rm I\hskip -2pt R}}

\def\e{\rm e}
\def\d{\delta}
\def\p{\partial}
 
\def\sqr#1#2{{\vcenter{\vbox{\hrule height.#2pt
\hbox{\vrule width.#2pt height#2pt \kern#2pt
\vrule width.#2pt}
\hrule height.#2pt}}}}

 \def\1/2{{\scriptstyle{1\over 2}}}
 \def\a/2{{\scriptstyle{3\over 2}}}
 \def\5/2{{\scriptstyle{5\over 2}}}
 \def\7/2{{\scriptstyle{7\over 2}}}
 \def\3/4{{\scriptstyle{3\over 4}}}

\font\steptwo=cmb10 scaled\magstep2

\magnification=\magstep1

\def\sqr#1#2{{\vcenter{\vbox{\hrule height.#2pt
\hbox{\vrule width.#2pt height#2pt \kern#2pt
\vrule width.#2pt}
\hrule height.#2pt}}}}

\def \r{\rightarrow}

\b

\def\picture #1 by #2 (#3){
  \vbox to #2{
    \hrule width #1 height 0pt depth 0pt
    \vfill
    \special{picture #3}  
    }
  }

\def\scaledpicture #1 by #2 (#3 scaled #4){{
  \dimen0=#1 \dimen1=#2
  \divide\dimen0 by 1000 \multiply\dimen0 by #4
  \divide\dimen1 by 1000 \multiply\dimen1 by #4
  \picture \dimen0 by \dimen1 (#3 scaled #4)}
  }

\parindent=1pc

 \vskip3cm
\smallskip

  {\ce {\steptwo   Heat and Gravitation. III. Mixtures
   }}

\b
  \ce{Christian Fr\o nsdal \footnote*{email: fronsdal@physics.ucla.edu}}
\b
  \ce{\it Physics Department, University of California, Los Angeles CA
  90095-1547 USA}
 \b

\def\sqr#1#2{{\vcenter{\vbox{\hrule height.#2pt
\hbox{\vrule width.#2pt height#2pt \kern#2pt
\vrule width.#2pt}
\hrule height.#2pt}}}}

\def \r{\rightarrow}

\b
  \no ABSTRACT ~    The standard treatment of relativistic thermodynamics 
  is not convenient for a systematic treatment of mixtures.
  It is proposed that a formulation of thermodynamics as an action principle
may be a suitable approach to adopt for a new investigation of that and other
problems in thermodynamics and astrophysics.

This third paper of the series applies the action principle to a  study
of mixtures of ideal gases. The
action for a mixture of ideal gases is, in the first approximation,  the
sum of the actions for the components. The entropy, in the
absence of gravity, is determined by the Gibbs-Dalton hypothesis.
Chemical reactions such as hydrogen dissociation are studied,
with results that include the Saha equation and that are more complete than
traditional treatments, especially so when gravitational effects are
included.  A mixture of two ideal gases is a system
with two degrees of freedom and should exhibit two kinds of
sound. This is not supported experimentally and the Gibbs-Dalton
lagrangian has to be modified accordingly.  In the presence of gravity the
Gibbs-Dalton hypothesis must be further modified to get results that agree with
observation.   The overall
conclusion is that experimental results serve to pin down the lagrangian
in a very efficient manner. It leads to a convenient theoretical
framework in which many dynamical problems can be studied, and in which the
incorporation of General Relativity is straightforward.

\vfill
PACS Keywords: Atmosphere, mixtures, action principle.

\ve
\no{\steptwo I. Introduction}

Most physical theories that have reached a high level of development are
formulated as action principles. This is true of General Relativity, but it is
not true of astrophysical applications of General Relativity. The reason for
this historical anomaly has been the need to incorporate thermodynamics, 
a highly developed theory that represents the exceptions: thermodynamics 
is not formulated as a dynamical action principle.

This is odd, since Gibbs' famous contributions to thermodynamics are
entirely based on axioms of minimum energy and maximum entropy, stopping just
short of introducing a lagrangian. That Gibbs' ``energy" can be regarded
as a dynamical hamiltonian, at least in certain contexts, is well known; see
Fetter and Walecka (1980).

That astrophysical problems, within a General Relativistic setting, are subject
to a treatment based on an action principle was conjectured independently by
Bardeen (1970), Schutz (1970) and Taub (1954). A concrete example, related to
the nonrelativistic theory of Fetter and Walecka,  was presented by us (Fronsdal
2007). But the absence of an established formulation of thermodynamics,
suitable for the extension to the domain of General Relativity, may have led to
a justifiable skepticism about the viability of this approach to astrophysical
problems.

Whence the present series of papers, intended to present a view of
thermodynamics that can be extended to the context of General Relativity,
much in the manner in which it was done in the last cited paper, but
with the thermodynamical interpretation more developed and better understood.

Our  approach to thermodynamics is characterized by  a complete dynamical 
framework based on  an action principle. The action principle that desribes adiabatic
configurations of an ideal gas was known. Here we expand the
context to include heat transfer, dissipation, mixtures and chemical
reactions.

This paper is preparation for taking up one more of the issues that, it
seems to us, cloud the traditional approach to relativistic
thermodynamics: the treatment of mixtures. Mixtures have more degrees of
freedom than those envisaged by Tolman's relativistic equations (Tolman
1934). We begin by  determining the   lagrangian for a mixturte of ideal gases. 
This requires experimental input and the first observation that is invoked is the
validity of the Gibbs-Dalton hypothesis concerning the entropy of mixtures.
When chemical reactions are included  we obtain an expression for the
degree of dissociation that is formally identical to the Saha equation.
Taking the latter as a convenient, approximate summary of experimental
results, we find that it places a very simple constraint on the entropy.
Once the lagrangian is determined all applications are straightforward.
The determination of the lagrangian  by means of observation is a main
strategy adopted in this paper.

A principal feature of an action principle is that all properties of a
system are strongly interrelated. Having arrived at a
lagrangian that is adequate for some applications,  one must try
to use the same lagrangian in other situations. And if other considerations
require a modification of  the lagrangian, then one has to accept that
applications already thought to have been done will have to be revisited.
For example, having found a way to account for some of the  observed features
of sound propagation in mixtures, we may find that difficulties arise when
the theory is applied to the problem of concentrations in the atmosphere.
Any modification of the lagrangian will have effects on all
applications. For another example, a
polytropic model of the atmosphere cannot remain polytropic when the energy of
the photon gas is taken into account,  as it must for very high temperatures.

\b

\ce{\bf Summary}
Section II is a brief account of the first paper in this series. The
principal dynamical variables for describing a simple gas are the
density, a velocity field and the temperature. Variation
of the action with respect to the temperature gives a fundamental
relation of the ideal gas, an expression for the entropy in terms of the
density and the temperature.  Any discussion of mixtures is intimately
concerned with entropy. A study of simple processes, including cooling
and free expansion,  reveals just enough of this concept to allow us to
understand mixtures, including chemical reactions.

To deal with  mixtures
we need additional input from experiment or from kinetic theory.  A key
statement that is taken from observation is the Gibbs-Dalton hypothesis about
the entropy of mixtures. Traditional applications make use of general
properties of the Gibbs function but all these properties are relative to the
Gibbs function of a certain ``reference configuration" and the investigation is
hampered by the fact that this latter function is not known. For this reason,
the theoretical treatment of reaction rates, for example,  has limited
predictive power and refers instead to a ``reaction constant" that is actually a
function of the temperature, to be determined by experiment. In
Section III we use the experimental information that is expressed by the
Gibbs-Dalton hypothesis   to determine the
lagrangian,  thus preparing the way for  many types of
applications.

Turning to chemical reactions, we find that other
experimental input is required. The first case studied, in Section III.3,  is
the simplest one, the atomization of a warm hydrogen gas.  The
Euler-Lagrange equations of motion lead directly to a formula for the
reaction rate in terms of a single entropy parameter. This equation is
formally identical to the simplest version of the Saha equation (Eggert
1919, Saha 1920).  Taking this equation as a convenient summary of
experimental results, we observe that this parameter is constant through
the region of dissociation. That is, observation makes an especially
simple statement about the lagrangian.
 
One could now,  for example, calculate the characteristics of
sound propagation in partly dissociated hydrogen. Other results that
follows immediately from the equations of motion are formulas for the
internal energy and the entropy. A lagrangian for more general chemical
reactions is conjectured, Section III.4.

In Section IV we use this lagrangian to study 
the propagation of sound in a mixture. One obtains a
system of two linked wave equations and two distinct normal modes, with
different propagation velocities. As far as we know, this phenomenon has
been recognized in the context of superfluid helium only
 (Putterman 1974).
But this prediction of dual sound speeds is contradicted by experiment, which
makes it necessary to modify the lagrangian that was built on the Gibbs Dalton
hypotheses, without necessarily giving up that hypothesis, since it applies to
equilibria only.

In Section V we introduce the gravitational field in the usual way,
by including the gravitational potential in the hamiltonian, in each of the
contexts considered and for which the lagrangian has been found.
Application of the Gibbs-Dalton hypothesis to this situation is ambiguous
and unsuccessful.
We again turn to measurements (in the earthly atmosphere) to determine
the entropy of a mixture in the presence of gravity, obtaining density
and temperature profiles in good agreement with observation.

In another paper in this series, we shall return to the context
that stimulated this research, the application of relativistic
thermodynamics to stellar structure and development. The generalization of
the simplest lagrangian to include the effect of a generally relativistic,
dynamical metric is straightforward (Fronsdal 2007).

As in all the papers of this series, viscosity is neglected
and the motion is irrotational. This limitation is not as strong as it may
appear to be; as may be seen from the existence of a hamiltonian formulation of
a theory of turbulence (Onsager 1949).

\b\b

 \no {\steptwo II. The ideal gas}\footnote*{The development presented here can
easily be expanded to apply, for example,  to a gas with an arbitrary, cubic
equation of state (Fronsdal 2010).}

\b

\ce
{\bf II.1. The extremal principles of Gibbs}

The basis for a close integration of the action principle with
thermodynamics is the identification of the hamiltonian with the function
$$
H = F(\V,T) + ST + P\V,\eqno(2.1)
$$
a function of 4 independent variables $\V,T,S$ and $P$: volume,
temperature, entropy and pressure. The function $F$ is the free energy.

The states of thermodynamic equilibrium are the points in this
4-dimensional manifold at which $H$ is extremal with respect to
variations of $T$ and $\V$,
$$
{\p H\over \p T}\Big|_{\V, S, P} = {\p F\over \p T}\Big|_\V + S =
0,\eqno(2.2)
$$
$$
{\p H\over \p \V}\Big|_{T, S, P}= {\p F\over \p \V}\big|_T +P =
0.\eqno(2.3)
$$
The system is defined by the choice of the free energy function $F$.
When this function has been specified the
two relations define a 2-dimensional surface in a 4-dimensional
manifold with coordinates $\V,T,S,P$.

In the case of an ideal gas,
$$
F(\V,T) = -\R T \ln (\V T^n),\eqno(2.4)
$$
and these relations give
$$
S/\R =  n+\ln{ ( \V T^n)},\eqno(2.5)
$$
and
$$
P\V = \R T,\eqno(2.6)
$$
respectively. The on shell internal energy  density is
$$
U =\big(F(\V,T) + ST\big)\Big|_{\rm on~ shell} = n\R T.\eqno(2.7)
$$
``On shell" means that the function is restricted to the surface defined
by (2.2). The last three relations define the ideal gas; we take that as
sufficient proof that (2.4) is a correct and useful expression for the
free energy, and that the expression (2.1) is a promising candidate to
serve as  hamiltonian.

The free energy is the only potential that is defined from
the beginning as a function of two natural variables. The internal energy
is a ``variable". It becomes a function of $\V,T$ and $S$ by
identification with $F + TS$, restricted to the $\V,T,S$ surface. It is
  regarded as a function of $\V$ and
$S$, defined by the Legendre transformation $U(\V,S) = F+ST$ and
elimination of $T$ by use of the relation (2.2).   The function $F$
occupies a special position.

There is a symmetry between energy and entropy, strongly emphasized by
Gibbs (1878). One can regard the entropy as a function of $\V, T, H$ and $P$,
$$
S = {1\over T}\big(H-F(\V,T) -P\V\big).\eqno(2.8)
$$
The conditions that this function be stationary with respect to
variation of the
variables $T$ and $\V$, with $H$ fixed, are exactly the same as (2.2-3),
$$
{\p S\over \p T}\Big|_{\V,U,P} = -{1\over T}\Big(S + {\p F\over \p
T}\Big|_\V \Big) = 0\eqno(2.9)
$$
and
$$
{\p S\over \p \V}\Big|_{T,U,P} =  -{1\over T}\Big({\p F\over \p
\V}\Big|_T  + P\Big) = 0.\eqno(2.10)
$$
Gibbs claims that the two statements, $\delta H = 0$ and $\delta S = 0$,
are equivalent. But that is not true in the local extension of the theory.

\b

\ce{\bf 2.  Local relations}

The local extrapolation of thermodynamics seeks to
promote these relations to field equations that describe local but stationary
configurations. The functions $F, H, S$ and $\V$ are given new interpretations
as specific densities and specific volume. The mass
density is
$\rho = 1/\V$ and densities $f, h, s$ are defined by
$$
f(\rho,T) = \rho F(\V,T),~~ h = \rho H,~~ s = \rho S.\eqno(2.11)
$$
The Hamiltonian is,
$$
\int_\Sigma\, d^3x \,h,~~ h =   \rho \vec v^2\,/2 +
\rho\phi + f(\rho,T) + sT  + P~.\eqno(2.12)
$$
The effect of an external gravitational field is represented by the
term $\rho\phi$.

 Variation with respect to $T$, with $\rho,S,T$ treated as
independent variables,  leads to
$$
{\p f\over \p T}\Big|_\rho + s = 0.\eqno(2.13)
$$
Variation with respect to $\rho$, with the mass and the volume fixed,
\footnote * {A complimentary variation of the density, with the mass
fixed but the volume not, gives the result that the internal pressure
$p$ must agree with the external pressure $P$ on the boundary. See Section
II.6. Any sub volume is in equilibrium with the pressure excerted 
by the gas  outside its boundary; hence $P$ is identified with the thermodynamic pressure $p$.} gives
$$
{h\over\rho} + \rho{\p (f/\rho)\over\p \rho}\Big|_T = \lambda. ~~(
\lambda  = {\rm constant}.)\eqno(2.14)
$$
The local thermodynamic  pressure is defined by
$$
 \rho{\p (f/\rho)\over\p \rho}\Big|_T- f = p ,
$$
then the last relation  (2.14) reduces to
$$
{\rm grad}{h+p\over \rho} = 0,\eqno(2.15)
$$
  In the case of the ideal gas
$$
f(\rho,T) = \R\rho T\ln{\rho\over T^n}, ~~  p = \R \rho T.
$$
The on shell value of $h$ is the internal energy density,
$$
h|_{\rm on~ shell} =  \rho \vec v\,^2/2 + \rho\phi + n\R \rho T
$$
and the variational equations reduce  to
$$
S = \R \ln(\V T^n),~~ {\rm  and ~~grad} \Big(\vec v\,^2/2 + \phi +
(n+1)\R T\Big) = 0.\eqno(2.16)
$$
At equilibrium ($\vec v$ = 0), in  the absence of external forces
($\phi = 0$), the
temperature is uniform.

\b

\no{\bf Remark.}  The Bernoulli equation, in the stationary case (all
fields time independent), is
$$
\rho\,{\rm grad}(\vec v\,^2/2) + \rho\,{\rm grad}
\, \phi +{\rm grad}\,\R\rho T = 0.
$$
It should be noted that it is  equivalent to (2.16) only if
the gas is isentropic, when $\rho^n/T$ is uniform (Fronsdal 2010).
\b
Reversing the roles of $h$ and $s$ we extremize the total entropy
$$
\int_\Sigma d^3 x \,s,~~ s ={1\over T}\Big( h - \rho \vec v\,^2/2 -
f(\rho,T)-P\Big).
$$
Variation with respect to $T$, with $\rho,T$ and $H = h/\rho$ treated as
independent variables,  gives the same result, ${\p f/ \p T} + s =0, $ but
variation with respect to $\rho$ gives an additional constraint,
$$
{s\over \rho}-{P-p\over \rho T} = {\rm constant}.
$$
Since $P = p$, this tells us that the
specific entropy of an isolated ideal gas is uniform. We have seen that
this additional constraint is needed to identify the thermodynamic
pressure $-\p F/\p \V$ with the gas pressure that appears in the Bernoulli
equation.

 The function that we referred to as the hamiltonian
actually deserves the name. In order to derive hamiltonian equations of motion we need
to identify canonically conjugate momenta, in particular, a variable
that is conjugate to the variable
$\rho$. As shown by Fetter and Walecka (1980), a velocity potential
 fills the role admirably, with $\vec v = -{\rm grad}\, \Phi$ by
definition. The continuity equation takes the form
$ \dot\rho = -\d H/ \d \Phi$, the Bernoulli equation (in integrated form)
is
$\dot\Phi =
\d H/\d \rho$.

\b\b

\ce{\bf II.3. Dynamical action principle}

The ideal gas is a system governed by an action
$$
A = \int dt\int_\Sigma d^3x \L[\rho,T,\Phi,k_0,P],
$$
where $\Sigma\subset \Rrm^3$ is connected, $k_0\in \Rrm, \rho,T$ and
$\Phi$ are scalar fields on $\Sigma$ and $P$ is a scalar field on the
boundary $\p\Sigma$ of $\Sigma$.

For the present, the domain $\Sigma$ is fixed and the external
pressure $P$ is irrelevant; the action is varied with respect to the
density $\rho$, the temperature $T$ and the velocity potential
$\Phi$. The irrotational velocity field $\vec v = -$grad $\Phi$ is
subject to the boundary condition
$$
\vec v|_{\p\Sigma} \quad{\rm is ~ tangential}.\eqno(2.17)
$$

The lagrangian   is
$$
\L= \rho\dot\Phi - h =  \rho(\dot\Phi-\vec v\,^2/2 - \phi +\lambda) -
V(\rho,T,S),
$$
where $\vec v = - $grad$~ \Phi$ and $\lambda$ is a lagrange multiplier
related to the fixing of the mass,
$$
M = \int_\Sigma d^3 x \rho,
$$
The potential $V$ is
$$
V(\rho,T,S) = f(\rho,T) + sT, ~~ S = s/\rho = -\R\ln k_0 = {\rm constant}.
$$

Variation of $T$ gives Eq.(2.13),
$$
\p V/\p T = 0,~~{\p f\over \p T} + s = 0.
$$
For the ideal gas it is
$$
\ln{\rho\over T^nk_0} = n
$$
Variation of $\Phi$ gives the equation of continuity (local conservation
law)
$$
\dot\rho + {\rm div}(\rho\vec v) = 0.
$$With the boundary condition (2.17) imposed on $\vec v$, it ensures
the global conservation of the mass.  Variation of $\rho$ leads to
$$
\dot\Phi - \vec v\,^2/2 -\phi + \lambda =   {\p V\over
\p \rho}.
$$
Taking the gradient we get, since the entropy is uniform,  the
differential form of Bernoulli's equation (Bernoulli 1738),
$$
\rho{D\vec v\over Dt}+\rho\,{\rm grad}\, \phi = -{\rm grad }~ p,
$$
with the gas pressure
$$
p := \rho{\p V\over \p \rho}\Big|_{T,S}-V. \eqno(2.18)
$$

Besides the mass, the only globally conserved quantity is the hamiltonian
$H = \int _\Sigma d^3x \,h$, with the density
$$
h = \rho(\vec v\,^2/2 + \phi) + V[\rho].
$$
The local conservation law is
$$
\dot  h + {\rm div}\big(\vec  v(h+p)\big) = 0.
$$
The parameter $k_0$ is free, independent of volume and mass.
That a free parameter
must appear is evident since it must be possible to change the
configuration of the system by
heat transfer.

The lagrangian, for each fixed choice of $k_0$, desribes configurations
of an ideal gas restricted to an adiabat.  The equations of motion
preserve  mass and energy and the system is thus in  a sense isolated.
It is already possible to discuss the ``changes" that form the central
subject of thermodynamics.

\b\b
\ce {\bf II.4. Cooling}

By ``cooling" is meant a slow loss of energy, as the system goes through
the equilibria of a sequence  of adiabatic lagrangians indexed by $k_0$,
this being the only parameter available.  It is regarded as a reversible
process, for it is supposed that heat can be supplied, slowly so as to
make the departure from equilibrium negligible, to reverse the cooling
process. The domain $\Sigma$ is supposed fixed, $\rho$ and $T$ uniform
and $\Phi = 0$; hence $\rho$ is fixed, unchanged during the process of
cooling.

During this process, the on shell internal energy density is
$u = n\R T\rho$, the total energy is $U = n\R T M$, and changes in $U,T$
and $k_0$ are related by
$$
dU = n\R M d T, ~dk_0 = d{\rho \over T^n} = -nk_0{dT\over T}.
$$
As this represents a heat loss, $\ln k_0$ is increasing.  Hence
$$
dU = -\R TM d\ln k_0 := TdS,
$$
with
$$
S := -\R M \ln k_0.\eqno(2.19)
$$
This relates the parameter $k_0$ to entropy. We  define the
entropy density by
$$
s = -\R \rho\ln k_0.
$$
The specific entropy density $s/\rho$ is uniform in this case.
From the point of view of
thermodynamics, with its primary emphasis on equilibria, the lagrangian
density is a function of the 3 independent variables $\rho, T$ and $k_0$.
The on shell condition (2.5) is the fundamental relation of the ideal gas:
$$
s = -\R\rho \,(\ln{\rho\over T^n}-n).
$$

A dynamical description of cooling as an ongoing process may rely on the
heat equation.

\b\b

\ce{\bf II.5. Free expansion}

It is a process that begins (at $t = 0$) with a configuration
in which the gas is uniformly distributed
in a fraction $\alpha$ of the volume, with density $\rho$ and
temperature $T$, at rest.  From these initial conditions the gas spreads,
eventually occupying the total volume. Thermodynamics deals mostly with
equilibria, but the adiabatic lagrangian allows for motion as well, and
it describes the agitated configurations of the expanding gas for $t>0$.
Since the equations of motion preserve energy,  this predicted, adiabatic
motion of the gas cannot lead to equilibria. To bring the gas to rest
another process is required, involving a change in the value of $k_0$.

It is observed that the gas comes to rest, and for this to happen a non
adiabatic process must take place, with a change in entropy.
In order that the time development described by the adiabatic lagrangian
retain its relevance, one must postulate that the additional process
take place on a much longer time scale.

We thus envisage an initial, rapid, adiabatic expansion, leading to
configurations of non zero kinetic energy, and a slow conversion of that
kinetic energy into heat. Without attempting to describe the latter, it is
nevertheless possible to predict the end result. If the energy of the
asymptotic, equilibrium configuration is the same as that of the initial
configuration, then since both are equilibria of the same ideal gas we have
$U = n\R TM = U' = n\R T'M$, the prime referring to the asymptotic state.
Hence $T = T'$, and since
$\rho' = \alpha\rho$,  $k'_0 = \alpha k_0$ and the change in the
function $S$ is
$
\delta S = -\R M \ln\alpha,
$
which represents an increase since the fraction $\alpha$ is less than 1.
In general, if some heat is
lost or supplied, we shall have
$$
\delta S := - \R M\delta \ln   k_0 = \R M\ln{k_0\over k'_{0}}.
$$
and on shell,
$$
\delta S = \R M \big(\ln {\rho\over \rho'} + n \ln{T'\over
T}\big).\eqno(2.20)
$$

\b\b

\ce{\bf II.6. Work}

We consider changes of volume, and begin with the case of a
cylinder with a piston.  The cross section is ${\cal A}$ and the
longitudinal variable is $z$, with $0<z< z_1$. Conjugate to $z_1$ is
the external pressure $P$ excerted by the piston.

A change $d\V = {\cal A} dz_1$ of the volume requires that the gas supply
energy in the form of work, $dE = Pd\V = P{\cal A} dz_1$. The total,
conserved energy of the gas is thus modified by a term $+P\V$ (an additive
constant has no significance) and the lagrangian is modified by the
inclusion of a term $-P\V = -P{\cal A}z_1$,
$$
\int_\Sigma d^3x \L \rightarrow   \int_\Sigma d^3x\Big(\rho\,(\dot \Phi -
\vec v\,^2/2 +\lambda) -
\R T\rho\ln{\rho\over T^nk_0}\Big)  - P{\cal A} z_1,
$$
The action is to be extremized, in particular, by variations that
vanish at $\p \Sigma$. The earlier equations of motion therefore continue
to hold. In addition, we must consider the one
parameter family of mass preserving variations of the form
$
\delta\rho/\rho = -\delta z_1/z_1, ~~\delta\Phi = \delta T = 0.
$
This gives an additional equation of motion,
$$
\int_{\p\Sigma}d^2 x \L - {1\over z_1}\int_\Sigma d^3 x\big(\rho(\dot \Phi
- \vec v\,^2/2 + \lambda) - \R T\rho(\ln{\rho\over T^nk_0}+1)\big)
 - P{\cal A} = 0.\eqno(2.21)
 $$
The first term comes from changing the boundary, the second from variation
of
$\rho$ and the third from the new term that was added to the lagrangian.
With $p = \R T\rho $ from the equations of motion; after
multiplication by $z_1 $ it is
$$
z_1\int_{\p \Sigma} d^2 x \L - \int_\Sigma d^3 x (\L-p) - P\V = 0.
$$
By the equations of motion, $\L = p$. The conclusion is that the
external pressure $P$ is equal to the average of the gas pressure over
the face
of the piston.

The hamiltonian associated with the expanded lagrangian,
$$
H = \int_{\p\Sigma} d^3 x\big(\rho \vec v\,^2/2 + \R T \rho\ln{\rho\over T^nk_0}\big) + P\V
$$
is identified, in the case of an equilibrium configuration, with the Helmholz free energy $U + P\V$.
In general, the extra term that is needed in the case of a variable
boundary can be expressed as
$$
\int_\Sigma d^3 x P,
$$
where $P$ is a field that coincides with the external pressure at
the boundary $\p\Sigma$.

In open atmospheres   $P = 0$ at the upper end; it  implies
  that $T = 0$ there.

\b
\no{\bf Remark.} In Einstein's theory of gravitation the cosmological
constant can be interpreted   as an
external pressure on space  time at infinity.

\ve

\no {\steptwo III. Mixtures}

\ce{\bf III.1. Basic postulates}

The adiabatic action for a system of two parts, occupying distinct
portions of a total domain $\Sigma$, with no interaction or mutual
constraint between them, is the sum of the adiabatic actions of each,
$$
\int _\Sigma\L = \int_{\Sigma_1}\L_1 d^3 z + \int_{\Sigma_2}\L_2 d^3 z,~~
\Sigma_1\cup\Sigma_2 = \Sigma.
$$
In the case of two ideal gases,
$$
\int_\Sigma\L = \sum\int_{\Sigma_i} d^3 x \Big(\rho_i(\dot\Phi_i-\vec v_i^2/2
-\phi +\lambda_i) - V_i\Big),
$$
with
$$
V_i = \R_i T\rho_i\ln{k_i\over k_{0i}}, ~~k_i = {\rho_i\over T^{n_i}},~~
i = 1,2..
$$
The temperature field is defined over $\Sigma$; we may write $T_i =
T\big|_{\Sigma_i}, ~ i = 1,2$.

We restrict ourselves, temporarily, to the approximation in which the
gravitational field is ignored. In this case there is a unique, static
solution of the equations of motion with uniform densities $\hat\rho_1,
\hat\rho_2$ and temperatures $\hat T_1, \hat T_2$ given by $\hat\rho_i =
M_i/\V_i$ and
$
(\e \hat  T_i)^{n_i} = {\hat\rho_i/k_{0i}}, ~~i = 1,2.
$

We want to know what happens to the system if at $t = 0$ the barriers
that have confined the two gases to their respective domains are removed.
Each gas will expand (diffuse) into the region originally
occupied by the other. This implies motion and kinetic energy. Since
energy is conserved, the adiabatic equations of motion do not lead to a
state of equilibrium. But a real physical system can be depended upon to
find its way to rest, it is therefore necessary to postulate some non
adiabatic process, with a longer time scale. As in the case of
simpler situations examined in Section II, and for the same reason, we
may suppose that this slow process takes the system through a sequence of
lagrangians indexed by $k_{01}$ and $k_{02}$, the only parameters that are
available.  The changes can be interpreted in
terms of a changing (increasing) entropy.

We expect, then, that the mixture will eventually be governed by the same
lagrangian (but both densities will extend to the whole domain),
with a new set $k'_{01}, k'_{02}$ of parameters. Can we predict these new
values?  More precisely,  the problem is to determine a relation between the two parameters.
As heat is added to or withdrawn from the mixtures the entropy changes, tracing a path through the
$k_{01}, k_{02}$ plane. We need to fix this path.
\vskip3cm
  \vskip1.1in
\def\picture #1 by #2 (#3){
  \vbox to #2{
    \hrule width #1 height 0pt depth 0pt
    \vfill
    \special{picture #3}  
    }
  }
\def\scaledpicture #1 by #2 (#3 scaled #4){{
  \dimen0=#1 \dimen1=#2
  \divide\dimen0 by 1000 \multiply\dimen0 by #4
  \divide\dimen1 by 1000 \multiply\dimen1 by #4
  \picture \dimen0 by \dimen1 (#3 scaled #4)}
  }

\parindent=1pc

\vskip-3cm
 
\epsfxsize.5\hsize
\centerline{\epsfbox{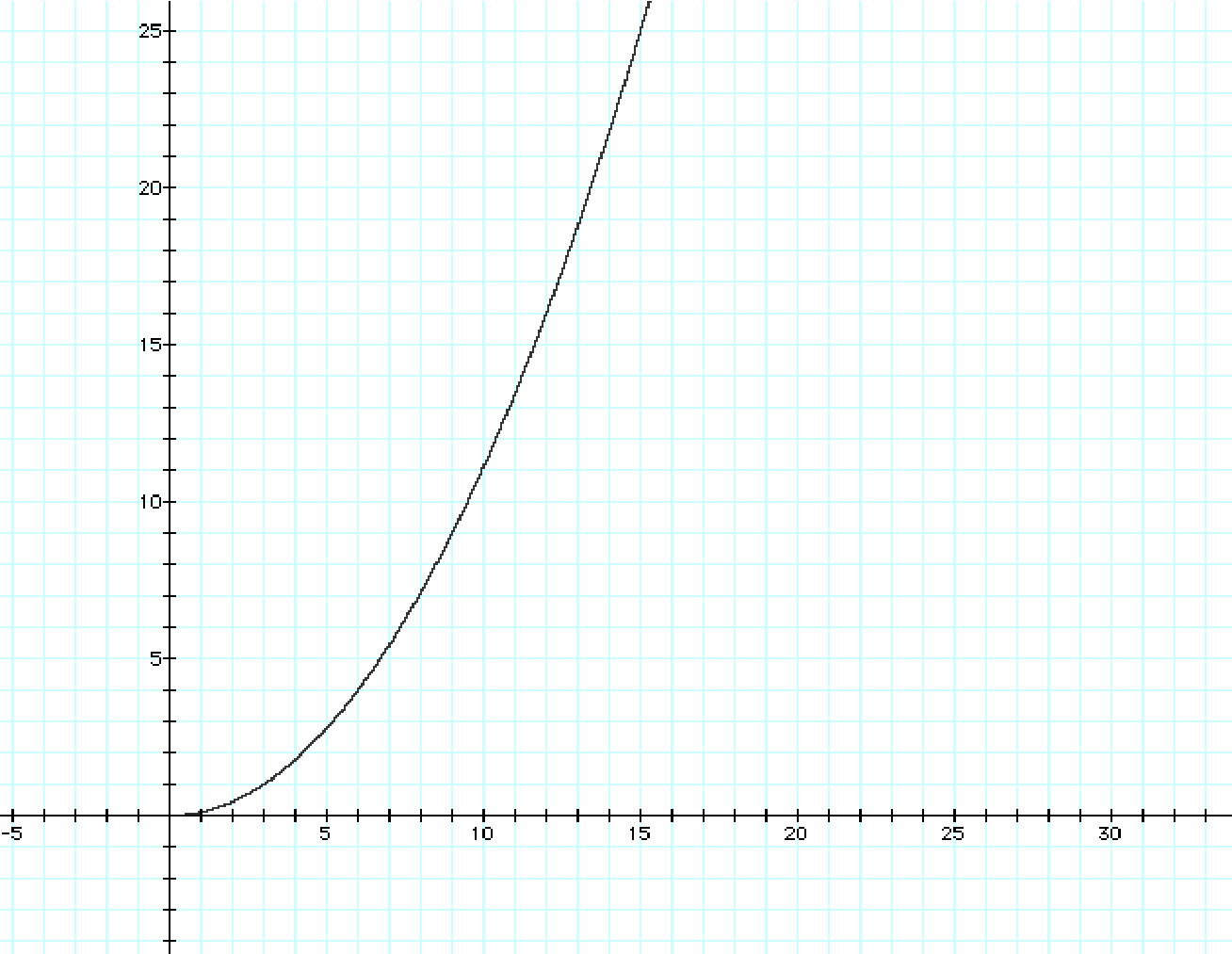}}
\vskip-.5cm

\b\b

Fig 1. Example of a path followed by a two component system in entropy space.
The coordinates are the entropy parameters $k_{01}$ and $k_{02}$ and the
shown shape conforms to the case of disassociation of atomic hydrogen.
\b

The Gibbs-Dalton hypothesis (Gibbs 1876,   Dalton 1808), justified
within the kinetic interpretation of the ideal gas as an ensemble of non
interacting particles, predicts that each gas undergoes \underbar{the same
change of specific} \underbar{entropy} as would be the case if the other gas
were absent. It is assumed that no chemical reaction is taking place.

The mathematical statement of the Gibbs-Dalton hypothesis is, in view of
Eq. (2.20)
$$
\ln{k'_{0i}\over k_{0i}} = \ln{\rho'_i\over \hat\rho_i} + n_i\ln{\hat
T_i\over T'},
$$
where $\hat \rho_i$ and $\hat T_i$ are the densities and temperatures
before mixing, $\rho'_i$ and $T'$ are the densities and the
common temperature
\underbar{of the final equilibrium configuration}.

The pre-mix parameters satisfy the relations
$$
\ln{\hat\rho_i\over \hat T_i^{n_i}  k_{0i}} = n_i,~~i
= 1,2.
$$
It may happen that $\hat T_1 = \hat T_2$, and we may wish to equalize the
temperatures before mixing. But in any case the
Gibbs-Dalton hypothesis is the condition that the two equations
$$
\ln{\rho_i'\over T'^{n_i}  k'_{0i}} = n_i,~~  i =
1,2, \eqno(3.1)
$$
with equilibrium values of the densities, are solved by the same value of
$T'$. A possible interpretation is that the difference between
the two temperatures defined by (3.1), dissipates with adjustment of the
entropies.

The extension of this constraint to non uniform configurations
is  problematic, as will be seen.
\ve

 This  Gibbs-Dalton constraint
imposes a relation between $k'_{01}$ and $k'_{02}$,
$$
(k'_{01}/M_1)^{n_2} = (k'_{02}/M_2)^{n_1}. \eqno(3.2)
$$
To the extent that the Gibbs-Dalton hypothesis is verified  experimentally
we can regard this constraint, and the lagrangian determined by it, as
being dictated by experiments. The lagrangian of the mixture
 is now determined by a single, real parameter that serves as a record of 
 heat that is lost or supplied.

A natural choice for the lagrangian is
$$
\L = \sum\Big(\rho_i(\dot\Phi_i - \vec v_i\,^2/2 -\phi +\lambda_i) - \R_i
T\rho_i\ln{\rho_i\over T^{n_i} k_{0i}}\Big),\eqno(3.3)
$$
with $\rho_1,\rho_2$ defined over $\Sigma$, the masses $M_i = \int_\Sigma
d^3 x \rho_i$ fixed, and constants $k_{0i}$ satisfying
$$
(k_{01}/M_1)^{n_2} = (k_{02}/M_2)^{n_1}.\eqno(3.4)
$$
We have dropped the primes on these parameters.

The equations of motion are the equations of continuity, the Bernoulli equation, namely
$$
\dot \Phi_i - \vec v_i\,^2/2 -\phi +\lambda_i = \R_i T(\ln
{k_i\over k_{0i}} + 1),~~ k_i := {\rho_i\over T^{n_i}}, ~~ i =1,2,
$$
and from variation of $T$ the adiabatic condition
$$
\sum\R_i \rho_i(\ln {k_i\over k_{0i}} - n_i) = 0. \eqno(3.5)
$$
Consider the uniform distribution, for which $\rho_i = M_i/\V$.
In that case,  because of
(3.4), both expressions $\ln {k_i/ k_{0i}} - n_i, i = 1,2, $ vanish for
one and the same value $T\,'$ of $T$.  We have
$ \ln {k_i/ k_{0i}}-n_i = n_i\ln(T\,'/T)$ so that the third equation
reduces to $\sum R_i\rho_in_i\ln(T\,'/T) = 0$, with the unique solution $T
= T\,'$.  Hence
$\ln {k_i/  k_{0i}} - n_i = 0,~ i = 1,2$, at equilibrium and both gases are polytropic.

Before mixing, both terms in (3.5) vanish separately on shell. The constraint (3.4) ensures that
there is a solution for which this is true for the mixture as well.

The total pressure; that is, the pressure on the walls, is
$$
P = \L-\sum \rho_i {\p\L\over \p \rho_i} = T\sum\R_i\rho_i.\eqno(3.6)
$$
Individual pressures are suggested by inspection of the
``Bernoulli equations" of motion,
$$
{\rm grad}\,( \vec v_i\,^2/2 + \phi + {\p V\over \p \rho_i}) =:~
{\rm grad}\,( \vec v_i\,^2/2 + \phi) + {1\over \rho_i} {\rm grad}\, p_i = 0.
$$
This yields $p_i = \R_i \rho_iT$ for equilibrium configurations, but
nothing that is useful in general. Additivity of pressures is valid in
the sense of (3.6), where each term is the pressure of a single gas
oppupying the total volume. It is also valid, but only at equilibrium, in
terms of the pressures that are identified through their role in the
Bernoulli equations.

\ve
\b\b

\ce{\bf III.2. About the Gibbs-Dalton hypothesis}

The constraint (3.4) is an expression of the Gibbs-Dalton hypothesis. This
hypothesis is widely used in the context of equilibrium thermodynamics.
It is, however, nonlocal. Consider a long tube filled with a mixture of two
ideal gases. One can wait until mixing is complete and
the temperature has become uniform over long distances. But if something
should perturb the system locally, even if the disturbance is uniform
over a moderate interval, then one is at a loss to understand how the
constraint, depending as it does on the total masses, is to be
implemented locally. We feel that a localized  version of the Gibbs-Dalton
hypothesis may be appropriate in such cases.

A  local version of the constraint would replace the masses by
the local densities; requiring that,  at equilibrium, both terms in
Eq.(3.5) vanish separately. But this would be inconsistent with the other
equations of motion whenever external forces, such as gravity,  are present.

The strongest version of the Gibbs-Dalton hypothesis would relate the two
densities to each other under all conditions, not just in equilibrium
configurations, thereby reducing the number of degrees of freedom.
 It seems that this is done routinely in some applications, which explains
the fact that dynamical processes have been described in terms of a single,
uniformly mixed density distribution. (Exception: superfluid helium.)

The truth may lie  somewhere between these extreme
interpretations of Gibbs-Dalton; between a local version
 that   admits
only one independent density degree of freedom, and the global version with
two densities that are locally independent while only the masses are
involved in the constraint. Implications for equilibrium configurations are
identical in all versions, in the absence of external forces.

One
way to think about this problem may be to postulate a slow, non adiabatic
process that is observed as a tendency of the mixture to anull the
difference
$
{\rho_1/ k_{01}}-{\rho_2/ k_{02}}.
$
It is an
irreversible process that calls for a role to be played by entropy.
A Lagrange multiplier may be interpreted as the limiting case of a force
that implements this tendency instantaneously. But it is characteristic of
all irreversible processes that they are slow, and they can be taken into
account in dynamical processes only under that condition. The local
version, with its strong identification of densities, is difficult to
justify.
  Only global versions  will be studied in this section.

\b\b

\ce{\bf III.3. Dissociation} We consider the simplest example of a
chemical reaction involving ideal gases, the transformation between
atomic and molecular hydrogen,
$$
H_2 \leftrightarrow 2H.
$$
  We fix, once and for all, the
domain $\Sigma$ with volume $\V$ and the total mass $M$.

At sufficiently low temperatures the gas is almost purely
molecular, with molecular weight and adiabatic index
$$
\mu_1 = 2,~~ n_1 = 5/2.
$$
At equilibrium we shall have $\vec v = 0$ and
$$
\rho_1 = \rho = M/\V, ~~ \ln{\rho\over T^{n_1}k_{01}} = n_1.
$$
At sufficiently high temperatures  the gas is almost purely
atomic, with
$
\mu_2 = 1,~~ n_2 = 3/2.
$
 At equilibrium,
$$
\rho_2 = \rho = M/\V,~~ \ln {\rho\over T^{n_2}k_{02}} = n_2.
$$

Consider a mixture of two gases, similar in all respects to $H$ and
$H_2$,
 except that the reaction analogous to $H_2 \leftrightarrow 2H_1$ does
not take place. If we assume the Gibbs-Dalton law for this mixture then
we are led to the lagrangian density
$$
\L =  \sum\Big(\rho_i(\dot\Phi_i - \vec v\,^2/2 -\phi +\lambda_i) -
\R_i
T\rho_i\ln{\rho_i\over T^{n_i} k_{0i}}\Big),
$$
with values of the parameters $k_{01}$ and $k_{02}$ that ensure the
compatibility of the on shell conditions
$\ln(\rho_i/T^{n_i}k_{0i}) = n_i, ~i = 1,2$ that must hold at
equilibrium. A hypothesis that can be tried is that this
Gibbs-Dalton lagrangian retain some validity for the real hydrogen
problem. But it must be changed in essential ways.

In the case of real hydrogen the individual masses are not conserved;
there is only one equation of continuity and only one velocity
field. Consider therefore
$$
\int\L = \int_\Sigma d^3 x   \Big(\rho(\dot\Phi - \vec v\,^2/2
-\phi  + \lambda)+\epsilon\rho_1 - \sum \R_i T\rho_i\ln{\rho_i\over T^{n_i}
k_{0i}}\Big),
$$
with $\rho = \rho_1 + \rho_2$ and $\epsilon>0$ constant.
This expression, with constant parameters $k_{0i}$, is a reasonable
candidate for hydrogen, but  the Gibbs-Dalton constraint
on the parameters is no longer applicable.

The term $\epsilon \rho_1$ is the  binding energy of molecular hydrogen.
Variation with respect to $T$ yields the  relation
$$
\sum \R_i\rho_i\big(\ln{\rho_i\over T^{n_i}k_{0i}}-n_i\big) =
0,\eqno(3.8)
$$
a relation that we shall put aside for now.

Variation of the densities with $\rho = \rho_1 + \rho_2$ fixed, in the
case of equilibrium, gives
$$
d\big(\sum \R_i\rho_i T \ln{\rho_i\over T^{n_i}k_{0i}}-
\epsilon\rho_1\big)
=0,
$$
or explicitly
$$
\R_1T\big(\ln{\rho_1\over T^{n_1}k_{01}}+1\big)
-\R_2T\big(\ln{\rho_2\over T^{n_2}k_{02}}+1\big) = \epsilon.\eqno(3.9)
$$
This relation appears in  textbooks on
thermodynamics, except that, in some of the books consulted  (Holman 1974,
DeHoff 1993),  the binding energy is not taken into account. And the
unknown entropy of a reference configuration used in these books is here
represented by the term $\sum\R_i T\rho_i \ln k_{0i}$ in
the lagrangian, parameterized by
$k_{01}, k_{02}$. Thes parameters, and thus the entropy of the reference state 
 not unknowable; but determined by the experiment.

Relations (3.8) and (3.9) are the equations of motion that determine
the configurations of adiabatic equilibrium, for each choice of the
parameters $k_{01}$ and $k_{02}$,  unique
equilibrium values of $T$ and for the ratio $\rho_1/\rho_2$. (The total
density $\rho = \rho_1 + \rho_2$ and the volume are both fixed.)

Addition or withdrawal of heat, or a change in energy   by other
means, produces a non adiabatic path through the 2-dimensional space of
lagrangians with coordinates $k_{01}, k_{02}$. To discover the relation
between $T$ and $\rho_1/\rho_2$ we must know this path. See Fig. 1.

Since $\R_2 = 2\R_1$, Eq.(3.9) reads
$$
\ln (q {\rho_1\over \rho_2^2}) - (n_1-2n_2)\ln T = \epsilon/\R_1T +1,~~ q
:= {k_{02}^2\over k_{01}},
$$
or

$$
{\rho_2^2\over\rho\rho_1}  = {r^2\over 1-r}  = {q\over\e \rho}
\,T^{1/2}\e^{-\epsilon/R_1T};~~ r:= \rho_2/\rho.\eqno(3.10)
$$

If the  parameter  $q$ is constant, then this is a special case of the
famous Saha formula (Eggert 1919, Saha 1920). That
$q$ is constant is thus another property of ideal gases that is
suggested by statistical mechanics. It is also supported by observation.
 The molecular
binding energy is about 4.5 ev, or in relation to the rest energy, in energy
units  $\epsilon = (4.5/1876 \times 10^6) c^2 \approx 2.15\times10^{12}$.
Thus
 $$
\epsilon/\R_1 = 4.3\times 10^{12}/.83214\times 10^8 = 51 674.
$$

Fig. 2 shows $r$ versus $T$, for $n_1 = 5/2, ~n_2 = 3/2$ with $q =
1$.   This result shows that a reasonable lagrangian for hydrogen across the
dissociation region is obtained by assigning a constant fixed value to
$q$.
\vskip3.5cm

\def\picture #1 by #2 (#3){
  \vbox to #2{
    \hrule width #1 height 0pt depth 0pt
    \vfill
    \special{picture #3}      }
  }
\def\scaledpicture #1 by #2 (#3 scaled #4){{
  \dimen0=#1 \dimen1=#2
  \divide\dimen0 by 1000 \multiply\dimen0 by #4
  \divide\dimen1 by 1000 \multiply\dimen1 by #4
  \picture \dimen0 by \dimen1 (#3 scaled #4)}
  }

\parindent=1pc

\vskip-3cm
 
\epsfxsize.6\hsize
\centerline{\epsfbox{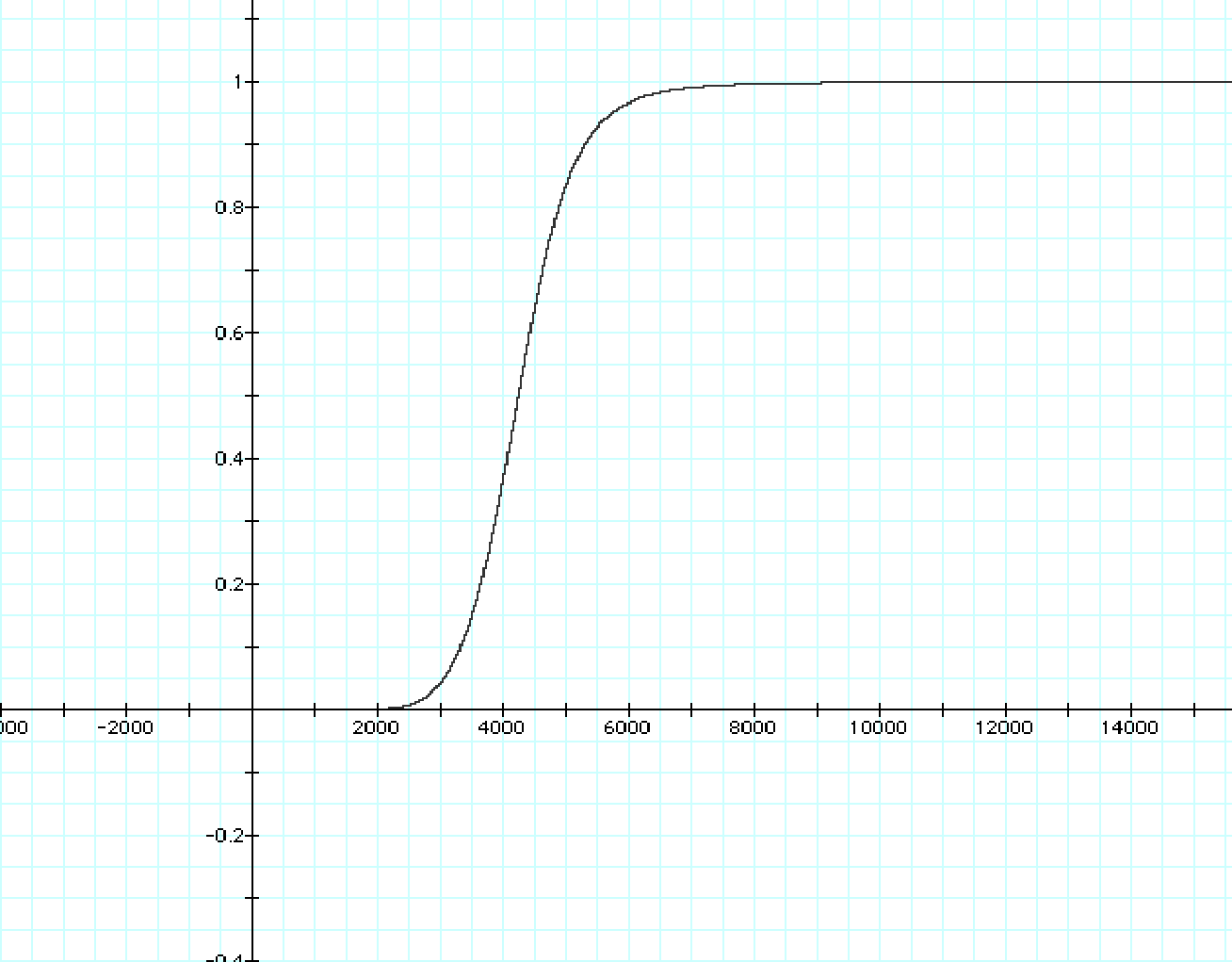}}
\vskip-1cm

\vskip1cm
 \ce{ Fig.2. The ratio $r$ against $T$, Eq.(3.10).}
\b
The measurements are usually performed under conditions of constant
pressure.
Setting $\rho_1/2 + \rho_2 = p/\R T$ we get, when $n_2 = 5/2$ and $n_2 =
3/2$,
$$
{\rho_2^2\over\rho\rho_1}p = {r^2\over 1-r^2}p = {\R q\over 2\e}
\,T^{3/2}\e^{-\epsilon/R_1T};~~ r= \rho_2/\rho.\eqno(3.11)
$$
This relationship is plotted in Fig. 3 for $q\R/p = 1$.

  \vskip1.5in
\def\picture #1 by #2 (#3){
  \vbox to #2{
    \hrule width #1 height 0pt depth 0pt
    \vfill
    \special{picture #3}      }
  }
\def\scaledpicture #1 by #2 (#3 scaled #4){{
  \dimen0=#1 \dimen1=#2
  \divide\dimen0 by 1000 \multiply\dimen0 by #4
  \divide\dimen1 by 1000 \multiply\dimen1 by #4
  \picture \dimen0 by \dimen1 (#3 scaled #4)}
  }

\parindent=1pc

\vskip-3cm
 
\epsfxsize.8\hsize
\centerline{\epsfbox{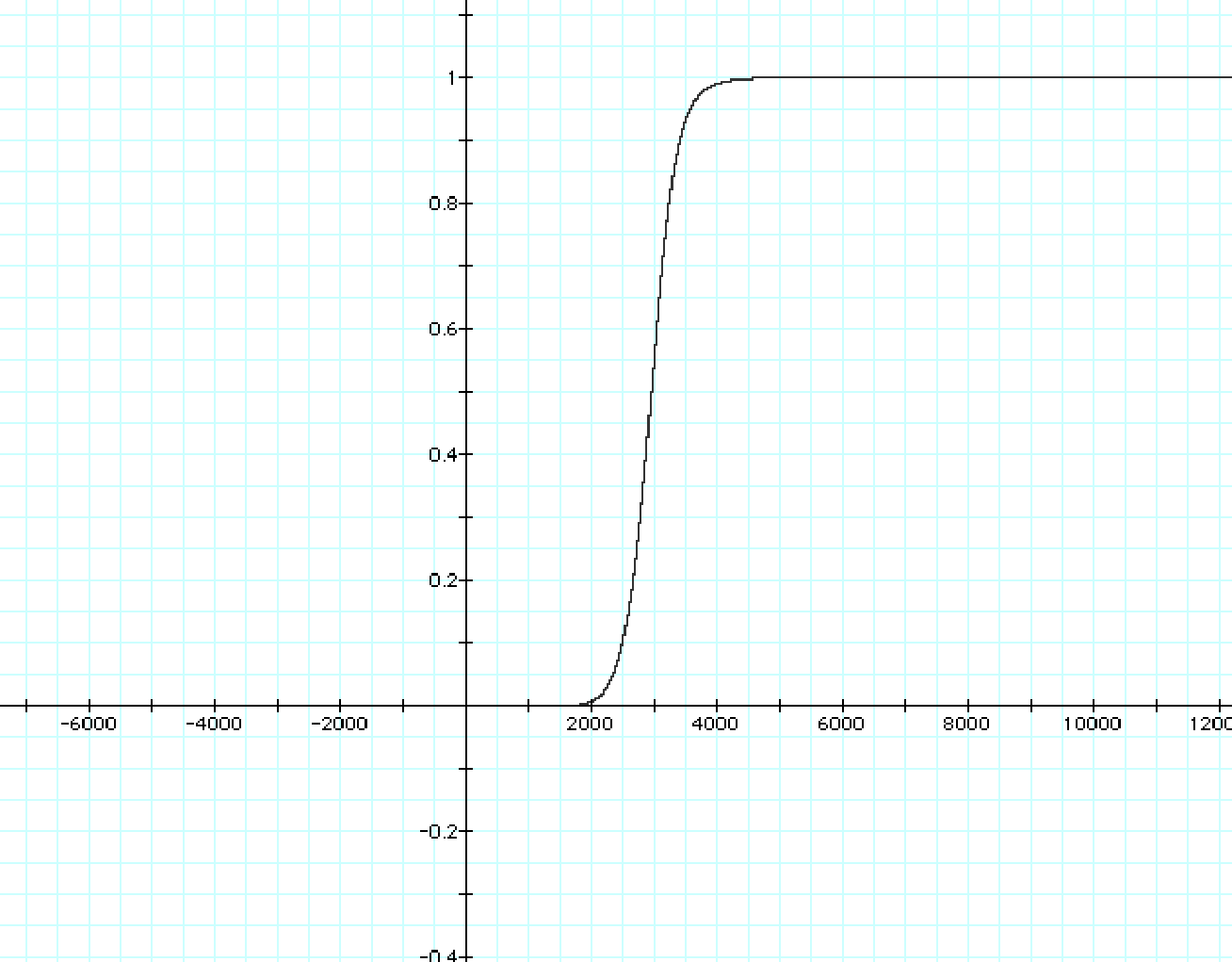}}
\vskip.5cm

\ce{Fig.3. The ratio $r$ against $T$, Eq. (3.11).}
\b

Eq.(3.10) applies to laboratory conditions of constant volume.
Eq.(3.11) applies under similar conditions when the gas is kept under
constant atmospheric pressure. Both equations can be used for atmospheres
if the dependence of temperature and density (resp. temperature and
pressure) on altitude are known.

Eq. (3.8)
  gives us other kinds of information. This
equation allows to calculate both $k_{01}$ and $k_{02}$ in terms of the
equilibrium values of
$\rho_1, \rho_2$ and $T$. With (3.10) it  allows to calculate all the
thermodynamic potentials as functions of the same variables. For example,
the expression for the on shell internal energy is the same as for the case
of a mixture, corrected only by the addition of $-\epsilon\rho_1.$

\b\b

\ce{\bf III.4. Other reactions}

We consider, briefly, a  chemical reaction involving 4 gases, by which
a uniform mixture of gases of types 1 and 2  are   transformed into
gases of types 3 and 4, according to the chemical equation
$$
\sum_{i = 1,2}\nu_i A_i  \leftrightarrow \sum_{i = 3,4}\nu_iA_i
$$
The coefficients are  the ratios of reagents, in grams.
Three masses are conserved, $\nu_2M_1 - \nu_1M_2, ~ \nu_3M_2 + \nu_2M_3,~~
\nu_4M_3-\nu_3M_4,$
 and the three corresponding densities satisfy conservation laws. The
kinetic part of the lagrangian  density therefore must take the form
$$\eqalign{
\L_{\rm kin} = (\nu_2\rho_1& -  \nu_1\rho_2)(\dot\Phi_1-\vec v_1\,^2/2 +
\lambda_1) \cr & +
 (\nu_3\rho_2 + \nu_2\rho_3)(\dot\Phi_2-\vec v_2\,^2/2 + \lambda_2)
+(\nu_4\rho_3-\nu_3\rho_4)(\dot\Phi_3-\vec v_3\,^2/2+\lambda_3)\cr}
$$
and
$$
\L = \L_{\rm kin} + \rho \phi +\epsilon\rho_1-\sum_{1\leq i\leq 4}\R_i
\rho_iT\ln{\rho_i\over T^{n_i}k_{0i}},
$$
where $\rho = \sum\rho_i$ and the term $\epsilon\rho_1$ represents the
energy needed to make the reaction go.

At equilibrium the densities are fixed by the volume and the masses, up
to a variation of the form
$$
(d\rho_1,...,d\rho_4) \propto(
\nu_1,\nu_2,-\nu_3,-\nu_4) .
$$
This, and the variation of $T$,  gives the two equations of motion
$$
\sum\R_i\rho_i(\ln{\rho_i\over T^{n_i}k_{0i}}+1) = 0,
$$
and
$$
\sum_{1\leq \leq4} {d\rho_i}\R_i(\ln{\rho_i\over
T^{n_i}k_{0i}}+1) =  d\rho_1\epsilon/\R_1 T.
$$
The latter  can be expressed as
$$
\ln{\rho_1^{\nu_1/\mu_1}  \rho_2^{\nu_2/\mu_2}\over
\rho_3^{\nu_3/\mu_3}\rho_4^{\nu_4/\mu_4}}  = \ln q + (n_1 + n_2 - n_3 - n_4)
\ln T -\epsilon/\R_1 T, ~~ q = {k_{01}^{\nu_1/\mu_1}  k_{02}^{\nu_2/\mu_2}\over
k_{03}^{\nu_3/\mu_3}k_{04}^{\nu_4/\mu_4}}
$$
one recognizes the law of mass action (Guldberg and Waage 1864, Gibbs 1875, Holman 1974, page 489), except that
here the right hand side is not an unknown function, but a function that is
known explicitly in terms of the 4 entropy parameters $k_{01},...,
k_{04}$.  This formula is formally the same  as one that one derives from
kinetic theory, it provides us with a hint that the parameter
$q $ may be an invariant.

It is tempting to speculate that $q$ is a constant, in that case
there remain 3 independent entropy parameters.  It seems likely that
they are subject to constraints of the Gibbs-Dalton type. Thus

$$
(k_{01}/M_1)^{n_2} = (k_{02}/M_2)^{n_1},~~
(k_{03}/M_3)^{n_4} = (k_{04}/M_4)^{n_3}.
$$
In that case the index of entropy is just one free parameter, as it should
be for a complete determination of the entropy.
\ve

\no{\steptwo IV. The propagation of sound}
\b

\ce {\bf IV.1. The ideal gas}

The dynamical equations that govern the adiabatic excitations of an ideal
gas
along a fixed direction are (Section II.3 and Laplace 1816)
$$
\dot\rho + (\rho v)' = 0,~~ \dot\Phi -\vec v\,^2/2 +\lambda = \R
T(\ln{\rho\over T^n k_0} + 1)\eqno(4.1)
$$
and
$$
\ln {k\over k_0} = n, ~~ k := {\rho\over T^n}.\eqno(4.2)
$$
For perturbations around static equilibrium, to first order, we have
$$
d\dot \rho = -v'\rho,~~ dk = 0,~~ \dot v = -\R(\ln {k\over k_0} +
1)dT';.\eqno(4.3)
$$
Thus
$$
d\ddot\rho/\rho = -\dot v' = \R(n+1) dT'',~~ dT'/T = nd\rho'/\rho,
$$
and finally
$$
d\ddot\rho/\rho = \R T (1 + {1\over n}) d\rho''/\rho.
$$
The speed of sound is thus $\sqrt{\R T \gamma}$, $\gamma := 1 + 1/n$.
Only one degree of freedom is excited.

Consider next the implications of the global version of the Gibbs-Dalton
constraint. In  a mixture of two ideal gases two degrees of freedom are
excited and two normal modes of propagation will appear.
  \vskip1.5in
\def\picture #1 by #2 (#3){
  \vbox to #2{
    \hrule width #1 height 0pt depth 0pt
    \vfill
    \special{picture #3}      }
  }
\def\scaledpicture #1 by #2 (#3 scaled #4){{
  \dimen0=#1 \dimen1=#2
  \divide\dimen0 by 1000 \multiply\dimen0 by #4
  \divide\dimen1 by 1000 \multiply\dimen1 by #4
  \picture \dimen0 by \dimen1 (#3 scaled #4)}
  }

\parindent=1pc

\vskip-3cm
 
\epsfxsize.8\hsize
 
\vskip1cm

\ce{\bf IV.2. Propagation of sound in Gibbs-Dalton  mixtures}

  The linearized equations for a
perturbation of a Gibbs-Dalton equilibrium configuration are
$$
d\ddot\rho_i = -\dot v_i' = \p_x^2{\p V\over \p\rho_i}\Big|_T,\eqno(4.4)
$$
$$
{dT''\over T} = {R_1 d\rho_1'' + R_2 d\rho_2''\over n_1\R_1\rho_1 +
n_2\R_2\rho_2}.\eqno(4.5)
$$
The second equation, - from Eq.(3.5) -  defines an adiabatic change.
It is assumed that the unperturbed densities and the unperturbed
temperature are uniform and that the perturbations depend on only one of
the spatial coordinates; the prime denotes the spatial derivative.

Let
$$
\kappa = \R_2/\R_1 = \mu_1/\mu_2;~~ \tau = \rho_1/\rho_2.
$$
then
$$
{dT''\over T} =   {d\rho_1\over \rho_1}{\tau\over A} +
  {d\rho_2''\over \rho_2}{\kappa\over A},~~ A:= n_1\tau +
n_2\kappa,\eqno(4.6)
$$
Next
$$
{d\ddot \rho_1\over\rho_1} = {\R  T\over \mu_1}{\tau\over A}
{d\rho_1''\over
\rho_1} + {\R T\over \mu_1}{\kappa\over A} {d\rho_2''\over
\rho_2} + {\R T\over \mu_1}{d\rho_1''\over \rho_1}.\eqno(4.7)
$$
$$
{d\ddot \rho_2\over\rho_2} = {\R   T\over \mu_2}{\kappa\over A}
  {d\rho_1''\over \rho_1} + {\R T\over \mu_2}{\tau\over A}
  {d\rho_1''\over \rho_1} + {\R T\over \mu_2}{d\rho_2''\over
\rho_2}.\eqno(4.8)
$$
The matrix of squared velocity is
$$
{\R T\over \mu_1}\pmatrix{1 + {\tau\over A} &
{\kappa\over A}\cr  {\kappa\tau\over A}&
\kappa + {\kappa^2\over A}\cr},~~~ A = n_1\tau + n_2\kappa.
$$

For mixtures we define the reduced speed $c$ in terms of the cgs speed $v
= c\sqrt{\R_1T}$. This reduced speed is a zero of the determinant of the
matrix
$$
\pmatrix{1 + {\tau\over A} -c^2&
{\kappa\over A}\cr  {\kappa\tau\over A}&
\kappa + {\kappa^2\over A}-c^2\cr} \eqno(4.9)
$$
The theory predicts  two modes, with speeds that vary with the
concentrations. Experimenters report a single mode. We shall see that this
discreapancy can be overcome, but it is of some interest to begin by describing
the predictions of the naive, Gibbs-Dalton model in some detail.

Two different types of binary mixtures must be described separately.
\b
\no{\it Type 1 mixture, ``similar" gases}

This is the case when $\kappa\gamma_2 >1$, for example:
\b

\ce{Nitrogen/Argon: $\kappa = 28/40,~ n_1 = 5/2, n_2 = 3/2,$}
\b
\no with $\kappa \gamma_2 =35/24$. There is a Laplace mode, with speed
intermediate between the adiabatic speeds of the two pure gases. The two
amplitudes are in phase. The dominant gas carries most of the
energy but the amplitudes are comparable.  A
second mode has a speed that interpolates between the Newton (isothermal) speeds
of the two pure gases. The two amplitudes are in opposite phase. The
amplitude of the dominant gas tends to xero in the limit when this gas is alone.

All this tends to be confirmed by experiment. The second mode seems to be
anti intuitive and there are several reasons why it may be expected to be
unobservable, or at least to justify the fact that it has escaped detection.
The prediction of the Gibbs-Dalton model needs only a minor correction to agree
with experiment. See Fig.4.
  \vskip1.1in
\def\picture #1 by #2 (#3){
  \vbox to #2{
    \hrule width #1 height 0pt depth 0pt
    \vfill
    \special{picture #3} 
    }
  }
\def\scaledpicture #1 by #2 (#3 scaled #4){{
  \dimen0=#1 \dimen1=#2
  \divide\dimen0 by 1000 \multiply\dimen0 by #4
  \divide\dimen1 by 1000 \multiply\dimen1 by #4
  \picture \dimen0 by \dimen1 (#3 scaled #4)}
  }

\parindent=1pc

\vskip-3cm
 
\epsfxsize.6\hsize
\centerline{\epsfbox{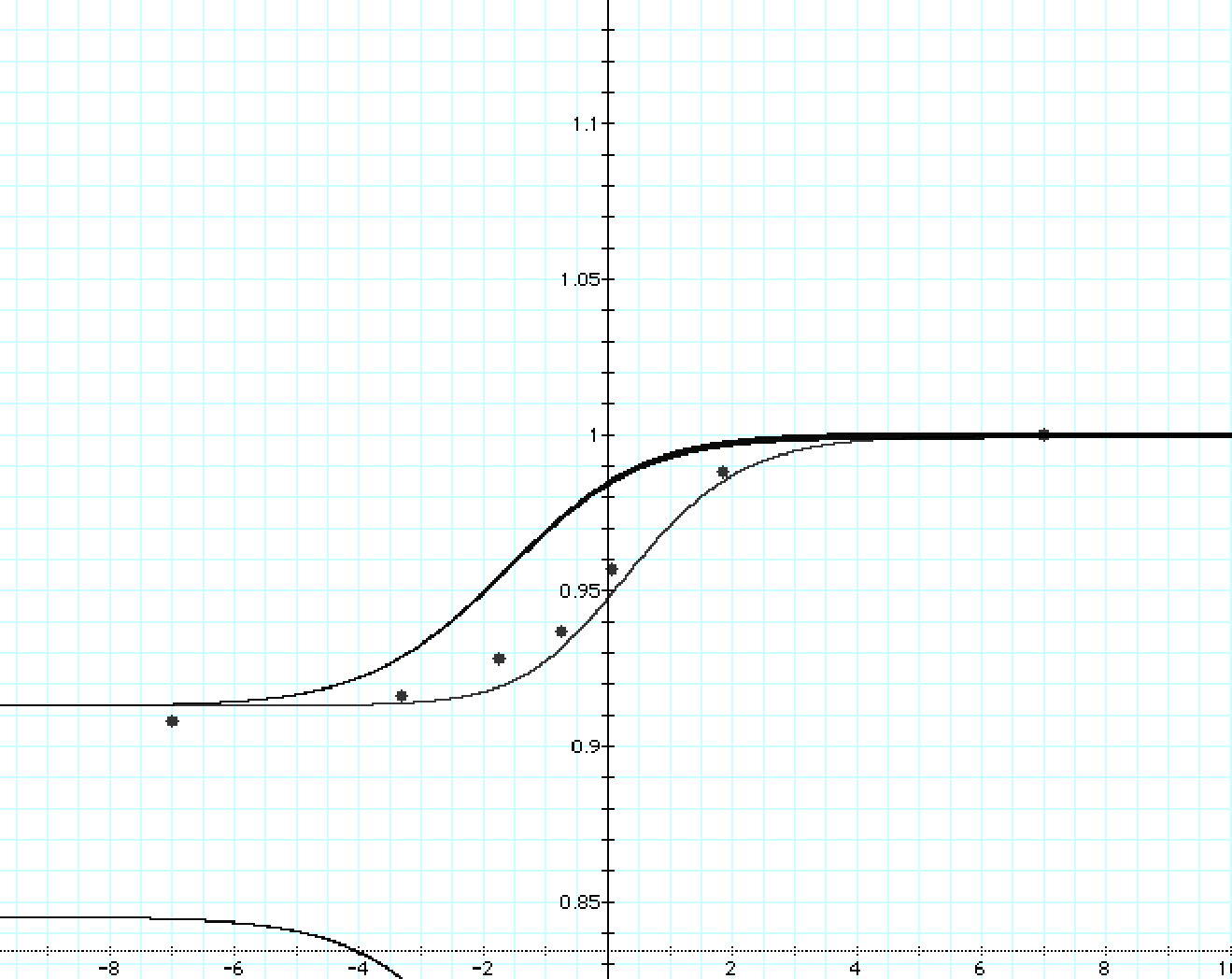}}
\vskip-1cm
\vskip1.2cm
 
   Fig 4. Speed of the two modes of sound in Ni/Ar,
 plotted against
ln${\rho_1\over
\rho_2}$. The curve just below
  the data points is the speed given by Eq.(4.10).
 
 \vskip.5cm
\no{\it Type 2 mixture, ``disparate" gases}

It is the case when $\kappa\gamma_2 < 1$, for example,

\ce{Helium/Argone, $\kappa = 4/40, n_1 = n_2 = 3/2$.}

\no There is a mode, with speed that approaches the adiabatic speed of sound in the
lighter gas in the limit when this gas is alone (NE corner), but in the opposite limit it
approaches the Newton value of the heavy gas (NW). The amplitudes are in phase.

A second mode has a speed that approaches the adiabatic speed of sound in the heavier gas
in the limit when this gas is alone (SW). Surprisingly, the amplitudes are in
opposite phase; the two gases are moving in opposite directions. See Fig.5a.

  \vskip.9in
\def\picture #1 by #2 (#3){
  \vbox to #2{
    \hrule width #1 height 0pt depth 0pt
    \vfill
    \special{picture #3}  
    }
  }
\def\scaledpicture #1 by #2 (#3 scaled #4){{
  \dimen0=#1 \dimen1=#2
  \divide\dimen0 by 1000 \multiply\dimen0 by #4
  \divide\dimen1 by 1000 \multiply\dimen1 by #4
  \picture \dimen0 by \dimen1 (#3 scaled #4)}
  }

\parindent=1pc

\vskip-2cm
 
\epsfxsize.5\hsize
\centerline{\epsfbox{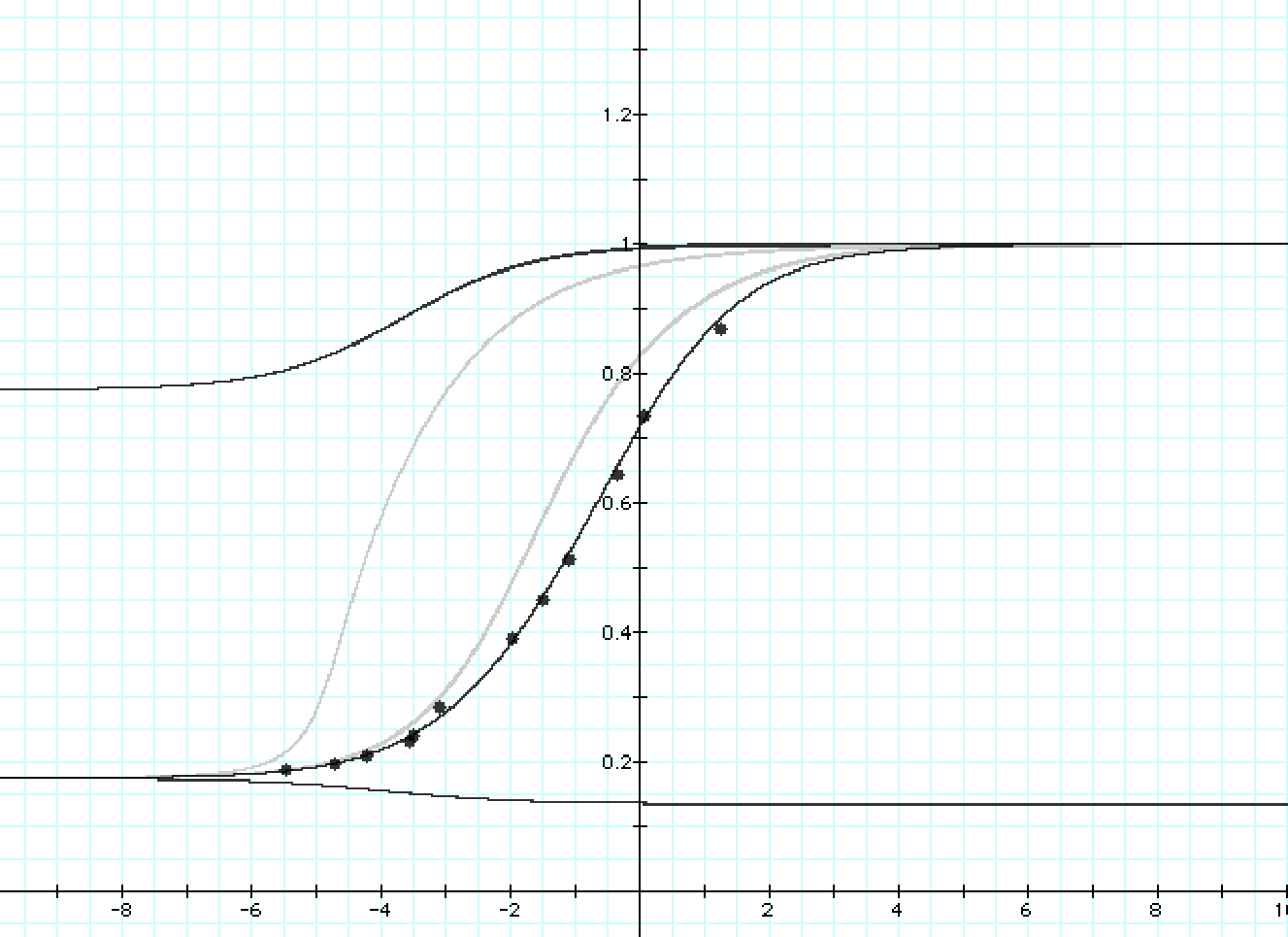}}
\vskip0cm

 Fig.5a. The speed of sound in $He/Xe$, a type 2 (disparate) mixture.
Coordinates as in Fig.4. The interpolating curve is from Eq.(4.10). Data points
from dela Mora (1986) and Bowler (.

\ve
\vskip1cm
  \vskip1.1in
\def\picture #1 by #2 (#3){
  \vbox to #2{
    \hrule width #1 height 0pt depth 0pt
    \vfill
    \special{picture #3}  
    }
  }
\def\scaledpicture #1 by #2 (#3 scaled #4){{
  \dimen0=#1 \dimen1=#2
  \divide\dimen0 by 1000 \multiply\dimen0 by #4
  \divide\dimen1 by 1000 \multiply\dimen1 by #4
  \picture \dimen0 by \dimen1 (#3 scaled #4)}
  }

\parindent=1pc

\vskip-3cm
 
\epsfxsize.6\hsize
\centerline{\epsfbox{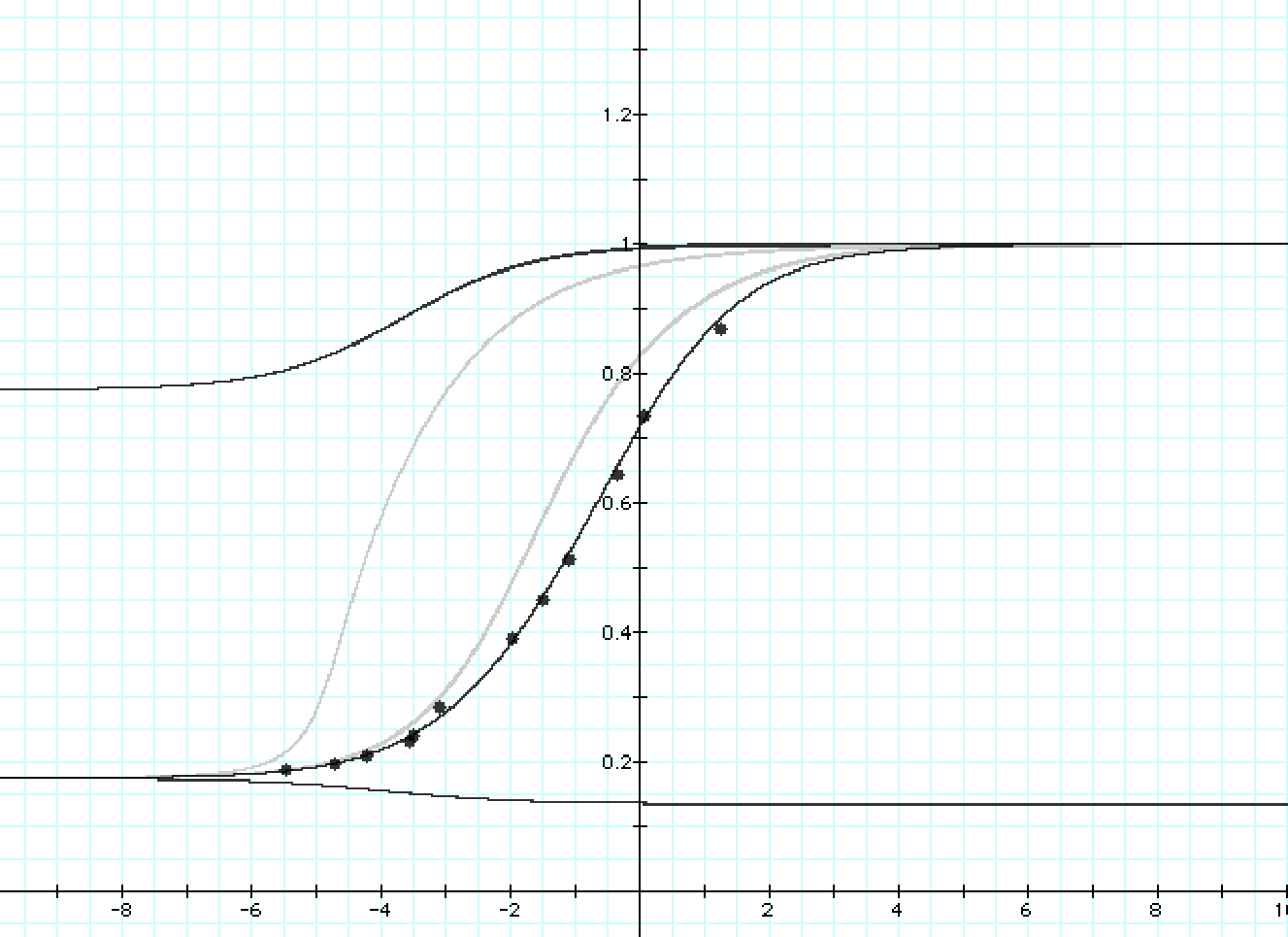}}
\vskip0cm

 Fig.5b. The speed of sound in $He/Ar$,
a type 2 (disparate) mixture.
Coordinates as in Fig.4. The
 interpolating curve is from Eq.(4.10).
The high datum point is
from Smorenburg 1996.
\vskip.5cm

The physical interpretation is more difficult for this type of mixture.
Experiment yields a single mode with a speed that varies with concentration
and that approaches the expected values in the limits of either pure gas.
It appears that a cross over takes place; in fact, there is evidence of a
``critical" concentration.

Experimenters report a single mode in all cases, with a speed
that varies smoothly with the concentration. There are frequent hints, however,
that there are other, less prominent modes, about which no information is given.

The reason for this disagreement was at first attributed to the fact that
no account had been taken of any kind of damping, and for high frequencies
this has been confirmed. But at low
frequencies the loss of energy to absorption is not thought to be
important. Indeed, the measured rates of absorption are extremely low.
(Holmes 1960)

The formula that fits the observations, and much better than could be
expected, is
$$
c^2 = {1\over \gamma_1}{\tau + \kappa\over \tau+1}\Big(1+ {\tau +\kappa\over
n_1\tau + n_2\kappa}\Big).\eqno(4.10)
$$
The origin of this prediction is not thermodynamics but calculations of
particle dynamics and the Boltzmann scattering equations. For a brief account
of the calculation see de la Mora and Puri (1986).   It is predicted to hold in
the limit when the effect of diffusion is so strong as to force the two
velocities to be nearly equal. That is the key to the success of the formula,
and that must be the basic fact that is responsible for its success: apparently
something forces the two velocities to be nearly equal. But the  conclusion
that it is due to strong diffusive damping is not compelling.

The equations that have been proposed to explain the success of (4.10)
are equations for the two velocities and, some times, two temperatures.
We have noted that the introduction of two temperatures may be interpreted
in terms of a deviation from Gibbs-Dalton equilibrium conditions. But
we are strongly constrained  by the need to retain the standard
equations, continuity and Bernoulli. Indeed, any modification of the
kinetic part of the lagrangian will cause a lack of conservation of
masses: that is, of particle numbers. It is therefore of interest to
explore alternatives.

The data quoted are from experiments at frequencies around 1 MHz, 
except for that of Smorenburg (...), the single, isolated  point at a high speed in Fig. 5b,
where the frequency was of the order of $10^9$ Hz. More data is needed, over a wide
range of frequency and temperature.

\b

 \b
\ce{\bf IV.3. A simple model of interactions}

Let us add  the following ``interaction term" to the potential; that is, to
the hamiltonian,
$$
\alpha \sqrt{\rho_1\rho_2},\eqno(4.11)
$$
with $\alpha$ constant. The reason for this choice will become clear almost
immediately. Such a term will affect the formula for the internal
energy, but it does not change the formula for the entropy. The
hypothesis of Gibbs and Dalton can be maintained, and we shall assume
values of the constants $k_{0i}$ accordingly. The formula (3.6) remains
valid; it is the Dalton rule of additivity of pressures.

The equations of motion that would be recognized as Bernoulli equations,
$$
{\rm grad}\,\Big( \vec v_i\,^2/2 + \phi + \R_i T\big(\ln{\rho_i\over
T^nk_{0i}}+1\big) + {\alpha\over 2}\sqrt{\rho_i\over \rho_j}\Big) = 0,
$$
no longer suggest a useful definition of individual pressures.
The acceleration of the molecules of each gas is strongly dependent on
the configuration of the other.

Again we consider first order perturbations of the equilibrium configuration.
 The linearized equations (4.4) are modified:
$$
d\ddot\rho_i = -\dot v_i' = \p_x^2\Big({\p V\over \p\rho_i}\Big|_T +
{\alpha\over 2}\sqrt{\rho_j\over\rho_i}\Big), ~~ j \neq i.
$$
Eq.s (4.7) and (4.8) gain additional terms,
$$
 {\alpha\over 4}\sqrt{\rho_2\over \rho_1}\Big({d\rho_2''\over\rho_2}-
{d\rho_1''\over\rho_1}\Big)~{\rm and} ~{\alpha\over 4}\sqrt{\rho_1\over
\rho_2}\Big({d\rho_1''\over\rho_1}- {d\rho_2''\over\rho_2}\Big).
$$
What makes the choice (4.11) special is the fact that, for a monocromatic wave,
these additions are proportional to the difference $v_1-v_2$ of the
velocities. Comparison with the standard theory shows that this is what is
needed.

The reduced speed of propagation is now obtained by setting to zero the
determinant
$$
\pmatrix{1 + {\tau\over A}-\beta \tau^{-\1/2} -c^2&
{\kappa\over A} + \beta \tau^{-\1/2} \cr  {\kappa\tau\over A} +\beta
\tau^{\1/2} &
\kappa + {\kappa^2\over A}-\beta \tau^{\1/2} -c^2\cr},~~ \beta :=
{\alpha\over 4\R_1 T}.
\eqno(4.12)
$$
In the limit of large $\beta$ the only eigenvector has $d\rho_1/\rho_1 =
d\rho_2/\rho_2$ and the  eigenvalue  $c^2$ is precisely as in (4.10).

Numerical results for a type 1 mixture are in Fig.6. It may be seen that a value
$\beta = .5$ is sufficient to bring the theory into substantial agreement with
the experient. Higher values of $\beta$ improves the agreement for low
frequencies but tends to spoil it when compared with the few data that are
available for the highest frequencies.

  \vskip1.5in
\def\picture #1 by #2 (#3){
  \vbox to #2{
    \hrule width #1 height 0pt depth 0pt
    \vfill
    \special{picture #3}  
    }
  }
\def\scaledpicture #1 by #2 (#3 scaled #4){{
  \dimen0=#1 \dimen1=#2
  \divide\dimen0 by 1000 \multiply\dimen0 by #4
  \divide\dimen1 by 1000 \multiply\dimen1 by #4
  \picture \dimen0 by \dimen1 (#3 scaled #4)}
  }

\parindent=1pc

\vskip-3cm
 
\epsfxsize.6\hsize
\centerline{\epsfbox{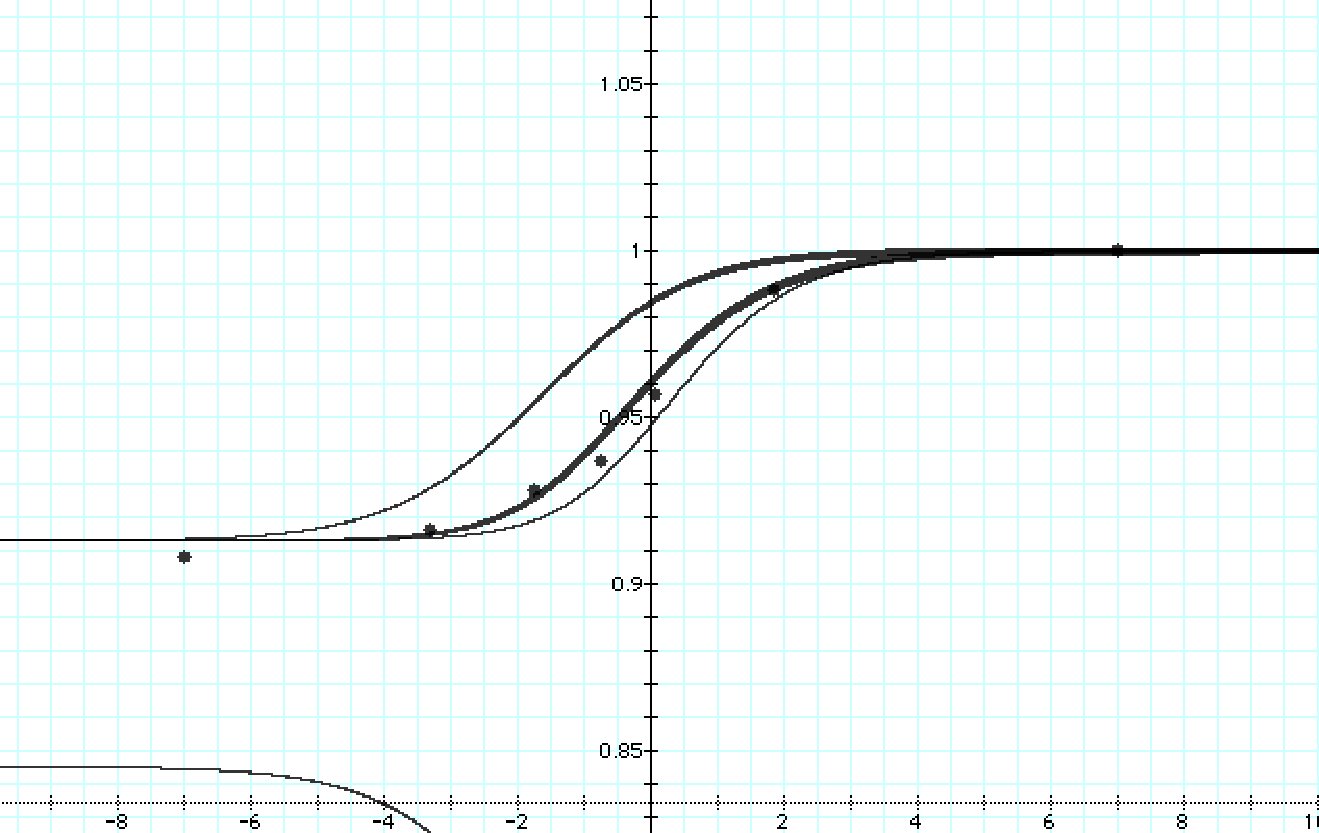}}
 
\b
 
  Fig.6. The effect of the interaction
 (4.11) on sound speed in $N_2/Ar$.
 The middle curve is from (4.12)
 with for $\beta = .5$.

\b
This modification of the lagrangian will be successful if the observation
of a single, dominant mode is limited to low and moderate temperatures. If
instead it extends to high temperatures an alternative may become more
attractive (see below). The theory predicts no dispersion.

The case of type 2 mixtures (Fig.7.) is more dramatic. very small values of
$\beta$ are enough to eliminate the isothermal modes in the SE and NW corners
of the figure.  The former disappears and the latter drops down to join the
Laplace branch at the lower left. Agreement with experiment requires a value of
$\beta$ at least equal to 1, near perfect agreement is gotten with $\beta = 5$,
and larger values of $\beta$ only improve the fit, except for the data at the
highest frequencies.

\vskip1.5cm
  \vskip1.1in
\def\picture #1 by #2 (#3){
  \vbox to #2{
    \hrule width #1 height 0pt depth 0pt
    \vfill
    \special{picture #3}  
    }
  }
\def\scaledpicture #1 by #2 (#3 scaled #4){{
  \dimen0=#1 \dimen1=#2
  \divide\dimen0 by 1000 \multiply\dimen0 by #4
  \divide\dimen1 by 1000 \multiply\dimen1 by #4
  \picture \dimen0 by \dimen1 (#3 scaled #4)}
  }

\parindent=1pc

\vskip-3cm
 
\epsfxsize.6\hsize
\centerline{\epsfbox{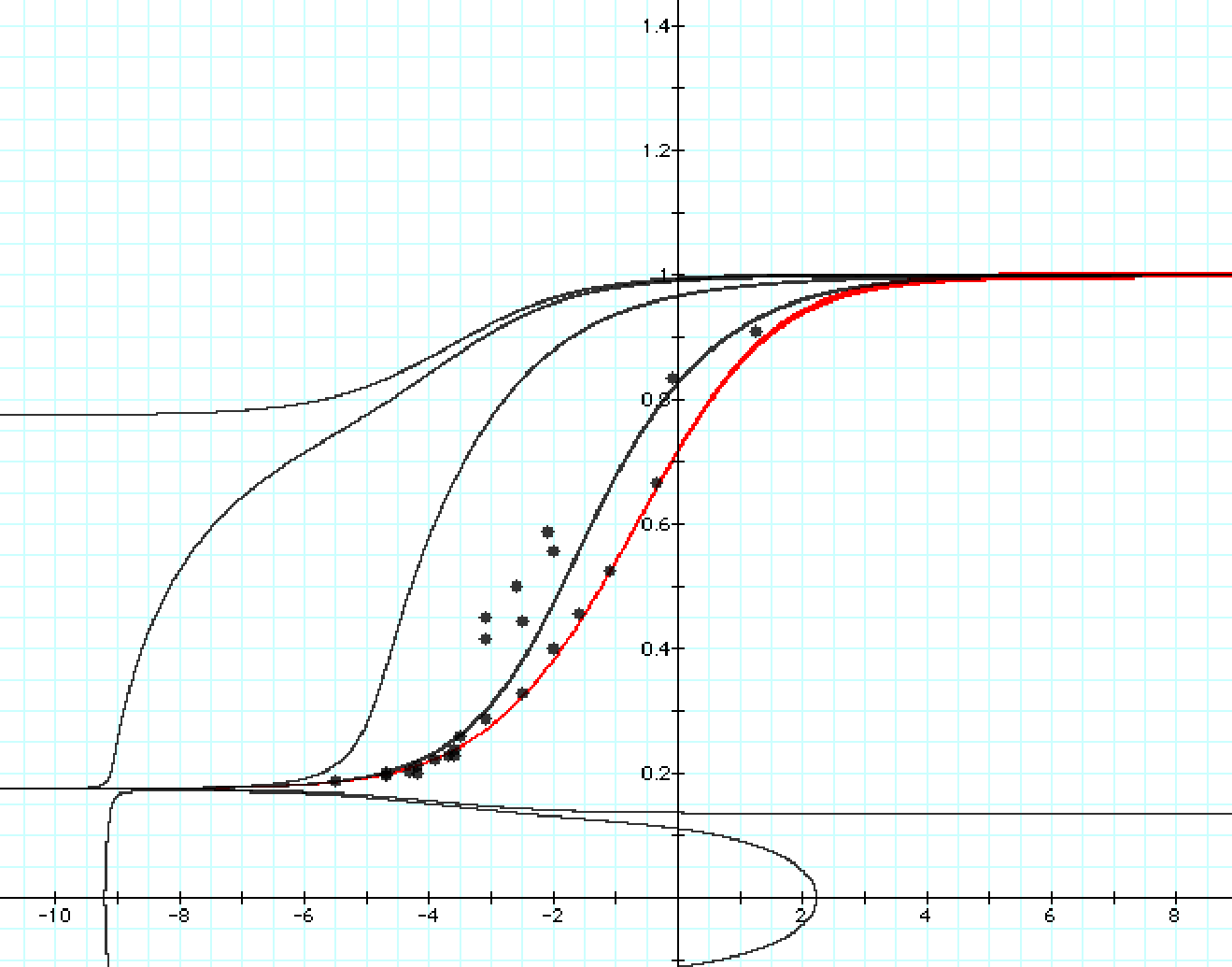}}
\vskip-1cm

\vskip1.5cm
  Fig.7. Effects of the simple interaction term on the speed of sound
in He/Xe, Eq. (4.12). From top to bottom: $\beta = 0, \beta = .01, \beta = .1,
\beta = 1;$\b\b

\ce{\bf IV.4. A variant}

One may try to try to replace $\alpha$ by $\alpha T$.
The potential is then
$$
V = \sum \R_i\rho_i T\ln{\rho_i\over T^{n_i}k_{0i}} + \alpha
T\sqrt{\rho_1\rho_2}.
$$
The new term may be considered as an addition to the free energy, and
 the expression for the entropy density is affected; it is possible
to retain the homogenous entropy predicted by  Gibbs-Dalton
theory. Thus we retain the former values of the parameters $k_{0i}$.

Variation of the temperature gives
$$
\sum\R_i\rho_i\ln{\rho_i\over (eT)^{n_i}k_{0i}} + \alpha\sqrt{\rho_1\rho_2} =
0.
$$
The individual gases are no longer polytropic, even at equilibrium.
The expression for the internal energy is independent of $\alpha$ and the
sum formula (2.6) for the  total pressure also remains valid.

Adiabatic changes obey
$$
{dT\over T} \sum R_i\rho_in_i = \sum \R_id\rho_i\Big( (\ln{\rho_i\over
(eT)^{n_i}k_{0i}} + 1\Big) +{\alpha\over 2}\sqrt{\rho_1\rho_2}\sum
{d\rho_i\over \rho_i}.
$$

Define
$$
N_i = \R_i\rho_i\ln{\rho_i\over (eT)^{n_i}k_{0i}},~~N = {N_1-N_2\over 2},
$$
then
$$
{dT\over T} \sum R_i\rho_in_i = \sum R_i d\rho_i + N
\Big({d\rho_1\over \rho_1}-{d\rho_2\over \rho_2}\Big).
$$

The linearized equation of motion is
$$
{d\ddot\rho_1\over \rho_1} = -\dot v_1' = \p_x^2{\p V\over \p \rho_1},
$$
with
$$
{\p V\over \p \rho_1} =  T\R_1(\ln{\rho_1\over T^{n_1}k_{01}} + 1) +
{\alpha\over 2}\sqrt{\rho_2\over \rho_1}.
$$
 The adiabatic variation is
$$\eqalign{
d{\p V\over \p \rho_1} &=   dT\big( R_1 + {N \over\rho_1}\big) +
\R_1T{d\rho_1\over \rho_1} + {\alpha\over 4}T\sqrt{\rho_2\over \rho_1}
\big({d\rho_2\over \rho_2} - {d\rho_1\over \rho_1}\big) \cr
& =  { T\over \sum\R_i\rho_in_i }\big( R_1 + {N \over
 \rho_1}\big)\Big(\sum R_i d\rho_i  + N\big({d\rho_1\over
\rho_1}-{d\rho_2\over \rho_2}\big)\Big) \cr &\hskip1in +\R_1T{d\rho_1\over
\rho_1}    + {\alpha\over 4}T\sqrt{\rho_2\over \rho_1}
\big({d\rho_2\over \rho_2} - {d\rho_1\over \rho_1}\big)\cr}
$$
This time, $c^2$ is an eigenvalue of the matrix
$$
\pmatrix{1+A_1( \rho_1 + N/\R_1)   - \beta\tau^{-1/2}   &
A_1\kappa(\rho_2-N/\R_2)   + \beta\tau^{-1/2}
\cr A_2( \rho_1+N/\R_1) + \beta\tau^{1/2} & \kappa +A_2\kappa(\rho_2 -N/\R_2)
-\beta\tau^{1/2}
\cr},
$$
where
$$
A_1 ={ 1\over \sum\R_i\rho_in_i }\big( R_1 + {N\over
\rho_1}\big)
$$
and $\beta = (\alpha/4\R_1)$.
For large $\alpha$ the eigen vector has $d\rho_1/\rho_1 = d\rho_2/\rho_2$ and
$\alpha$ is eliminated by adding $\rho_1$ times the sum of the elements in the
first row and
$\rho_2$ times the sum of the elements in the second row, with the result
$$
A_1\R_1\rho_1^2 + \R_1 T\rho_1 + A_2\R_2\rho_2^2 + \R_2 T \rho_2-v^2(\rho_1
+\rho_2) =0,
$$
which again agrees with (3.10).

Both models can account for the experiments within a range of temperatures
and frequencies. The only observation at a temperature other that room
temperature that is known to us is that of Smorenburg et al (Smorenburg 1995).
At a molar concentration of 3:1 in He/Ar (Type 2), ($\rho_1/\rho_2 = .3$)
, at 160K and 370 bar,
they observe the same sound speed as in pure helium. This may indicate that,
of our two models, our second one is closer to observation.

\ve

\no{\steptwo V. Application to atmospheres}
\b
\ce{\bf  V.1. Introducing gravitation}

Gravitation enters as a component of most, if not all applications of
physics. There
is a set and tested procedure for introducing the gravitational field
into any dynamical context. In classsical, non relativistic physics it
consists of adding the gravitational potential energy to the hamiltonian.
For an ideal gas one adds
$$
\int_\Sigma d^3 x\, \rho\,\phi,
$$
where $\rho$ is the mass density and $\phi$ is the Newtonian potential.
The adiabatic lagrangian becomes
$$
\int\L = \int_\Sigma d^3 x \Big(\rho(\dot\Phi-\vec v\,^2/2 -\phi
+\lambda) - \R T\rho\ln{k\over k_0}\Big).
$$
The gravitational field appears in the equation of motion,
$$
\dot\Phi - \vec v\,^2/2 -\phi +\lambda = \R T(\ln{k\over k_0} + 1),~~ k =
\rho/T^n.
$$ When the effect of gravity is taken into account in the dynamics
we get a theory of atmospheres. In the case of terrestrial gravity $\phi
= gz$, where $g$ is a constant and $z$ is the elevation. At equilibrium
$k$ is constant and the equation
$$
\lambda - gz = \R T(n+1)
$$
\no predicts a constant temperature gradient (lapse rate). The effect
has never, to our knowledge, been tested in the laboratory (but see Graeff
2009), but the constant lapse rate is a feature that is observed in the
earth's atmosphere and in the internal structure of stars (Lane 1870,
Ritter 1880, Eddington 1926). For the earthly troposphere this formula 
actually gives a value of the temperature gradient that is closed to 
observation.

We have noted that this successful application of  a theory originally
constructed to account for laboratory experiments where gravitational
effects are insignificant, modified in standard fashion to include
gravity, does \underbar {not} attribute the observed temperature gradient
to the radiation from the sun or to any external source other than
gravity (Fronsdal 2010).

 And yet it is evident that, in the absence of the sun, the present state
of the earthly atmosphere could not endure. The extinction of the sun
would lead to a general cooling of the atmosphere. This cooling is not
described by any process encompassed by the adiabatic lagrangian
dynamics. Instead, an external agent enters the picture, the spontaneous
emission of infrared radiation. As this is a slow effect we can regard the
cooling as a sequence of equilibria of adiabatic dynamics, a slow loss  of
 entropy and an increase in the value of $k_0$. (Section II.4.) The reverse
effect is also possible and if the terrestrial atmosphere is stable over long
times then we must conclude that spontaneous cooling by emission is balanced by
heating provided by the sun.

  \ve

\ce{\bf V.2. Mixed atmospheres}

At one time it was believed that, in an atmosphere consisting of several
components with different molecular weights, the lighter gas would float
on top. This was surely based on observation, since cooking gas,
entered at ground level, tends to remain there; however, it does so only
for a short time. Dalton   made the radical proposal that each gas
behaves as if the other were absent. This was an overstatement
and  led to much misunderstanding and debate, some of it recorded in
Dalton's book (Dalton 1806). In the case of non evanescent gases that have
had time to settle, Dalton's prediction is fairly close  to the truth, but
he did not go far enough, for actually it is observed that the
concentrations tend to  be  independent of elevation.
In order to account for this,
  one needs to invoke the intervention of a
mechanism, called  convection for short, or ``mixing due to fluid motions",
or diffusion.

This takes us back, once more, to consideration of a dissipative process,
which can be interesting, but if the problem that concerns us is the
density and temperature profiles  of the mixed atmosphere then it begs the
question. If, as is often observed, an equilibrium is ultimately reached,
then we are mainly interested in the end result, and less in
describing the process that leads to it. The process is dissipative and
entropy producing; its principal feature is that it is slow. It can be
ignored in the study  of the final equilibrium configuration, and the adiabatic
perturbations of it.

We have embraced the Gibbs-Dalton hypothesis and we have incorporated it
into our lagrangian for the mixtures that are  not under the influence of
gravity.   But in the presence of gravity there is an ambiguity. We shall
describe an attempt to adopt the global version of the hypothesis to
atmospheres. It is  only moderately successful. Then we shall determine the
lagrangian by direct  appeal to experience, to get a perfect fit.

\b\b

\ce{\bf  V.3. The Gibbs-Dalton model}

Consider at first a vertical column of air consisting of nitrogen and oxygen
and confined to the range $0<z<z_1$
of elevation. Because the Lagrange multipliers are freely adjustable we can
fix $z = 0$ at ground level. If gravity were not present we should have
an equilibrium configuration with uniform densities and temperature and
the equations of motion would give us
$$
\rho_i  = k_{0i}(\e T)^{n_i},~~\lambda_i = \R_iT(n_i+1),~~ i = 1,2,
$$
implying  the Gibbs-Dalton constraint, at equilibrium
$
\big({\rho_1/k_{01}}\big)^{1/n_1} = \big({\rho_2/
k_{02}}\big)^{1/n_2}
$
and thus
$
\big({M_1/k_{01}}\big)^{1/n_1} = \big({M_2/
k_{02}}\big)^{1/n_2}
$

Having chosen the domain and the parameters (except the Lagrange
multipliers), we add the gravitational field. The masses, presumed
unchanged,  are now related to the partial pressures at ground level,
\footnote*{It is assumed that the column extends upwards until the pressure
(and the temperature) becomes zero.}
$$
gM_i  = \R_i \rho_i(0)T(0). ~~ i = 1,2.
$$
The relative concentration of the heavy gas, $\rho_2/\rho_1 =
(\mu_2/\mu_1)(M_2/M_1)$ is increased at the lower level. We take the
total density at ground level to be .0012$, the temperature 300K,
the mass ratio $3.26:1  ($\rho_1(0)$ = $3.26 \rho_2(0)$) and
$
\mu_1 = 28,~~ \mu_2 = 32,~ \rho_1(0)= .9184\times 10^{-3},~~\rho_2(0) =
.2816\times 10^{-3},~~   n_1 = n_2 = 2.5.
$

The constraint (3.4) gives, for the masses implied by these parameter
values,
$$
k_{02}/k_{01} = ( \mu_1/\mu_2)(1/4) = .26825.\eqno(5.1)
$$
 The equations that determine the equilibrium configurations are
$$
\lambda_1 -\phi = \R_1 T(\ln x + 1 + n_1),~~ x := {\rho_1\over ({\rm
e}T)^{n_1} k_{01}},\eqno(5.2)
$$
$$
\lambda_2 -\phi = \R_2 T(\ln y + 1 + n_2),~~ y := {\rho_2\over ({\rm
e}T)^{n_2} k_{02}},\eqno(5.3)
$$
$$
\R_1\rho_1\ln x + \R_2\rho_2 \ln y = 0.\eqno(5.4)
$$
It is useful to note that
$
 {y(0)/ x(0)}  = {\mu_2/ \mu_1}.
$
The third equation can be rearranged to give, since $n_1 = n_2$,
$$
\ln x(0) = {1\over1 + \mu_2\rho_1/\mu_1\rho_2}\ln{\mu_1\over \mu_2},~~
\ln y(0) = {1\over1 + \mu_1\rho_2/\mu_2\rho_1}\ln{\mu_2\over \mu_1}.
$$
Thus
$
\ln x(0)= -.02824,~~ x = .9722,~~ \ln y =.1053,~~ y = 1.111
$
and
$$
k_{01} =  \rho_1(0)/(\e T)^nx = 4.975\times 10^{-11},~ k_{02}=
\rho_2(0)/(\e T)^ny =1.339\times 10^{-11}\eqno(5.5)
$$
and finally
$
\lambda_1 = 3.09645\times 10^9,~~ \lambda_2 = 2.81343 \times 10^9.
$

 The atmosphere ends where the pressure vanishes; that is, where $T =
0$.  Since $\lambda_1 > \lambda_2$,  this happens at the point
where
$$
\phi= \lambda_1, ~~ x = 1,~~ y = 0,	~~		 z = \lambda_1/g \approx 31.6\times
10^5~ (32~ km).
$$
 Just below this
point we find that
$
dT\propto dz,~~ \rho_1 \propto T^{n_1},~~ \rho_2 \propto T^{n_2}{\rm
e}^{-\alpha/T},
$
with $dz = \lambda_1/g-z$ and $ \alpha = (\lambda_1-\lambda_2)/\R_2
= 121.885.$
The first gas is dominant and adiabatic as if it alone constituted the
atmosphere; the second gas has a  different profile in this region.

More interesting are the temperature and density profiles near the
ground; we have
$$
-{T'\over T}(0) = 3.2303\times 10^{-7} =3.23~\%/km.\eqno(5.6)
$$
In a nitrogen (oxygen)  atmosphere the numbers are 3.14\% (resp 3.59 \%).
The density profiles are
$$
{\rho_1'\over\rho_1}(0) = -7.852\times 10^{-6},~~ {\rho_2'\over\rho_2}(0) =
-9.038\times 10^{-6}.\eqno(5.7)
$$
The last two ratios are found by observation to be very much closer. The
concentrations are reported with four significant digits (no error bars) at
all levels where the theory can be applied with any confidence (up to 10
km).

The slopes were calculated as follows. From (5.2-3),
$$
-g\mu_1 = \R T{\rho'_1\over \rho_1} + \R T'(\ln x + 1),~~
-g\mu_2 = \R T{\rho'_2\over \rho_2} + \R T'(\ln y + 1)
$$
From (5.4) and some reduction,
$$
{T'\over T} = {-g\mu_1\mu_2\over \R T}{\rho_1(\ln x + 1) + \rho_2(\ln y + 1)
\over \mu_1\rho_2[(\ln y+1)^2 + n_2] + \mu_2\rho_1[(\ln x + 1)^2 + n_1]} .
$$

The prediction (5.7), of two different rates  of decrease of density, is not
very far off. It may be correct when the  rate of diffusion, that tends to
reduce the difference,  is small, and  for young mixtures, when diffusion did
not yet have enough time to act.  The difference is only 1 percent per km, but
it is much too large to be reconciled with measurements in our atmosphere.
\b\b

\ce{\bf V.4. Entropy from experiment}
Let us take it as a result of observation that the concentration of the
principal gases in our atmosphere are very nearly constant at low
altitudes, and use this datum to find the values  of the entropy parameters
$k_{01}, k_{02}$.

The 3 equations of motion are
$$
\sum R_i\rho_i \ln x_i = 0,~~ x_i := {\rho_i\over k_{0i}
(eT)^{n_i}},\eqno(5.8)
$$
$$
\lambda_i -gz = \R_i T (\ln x_i + 1 + n_i),~~ i = 1,2.\eqno(5.9)
$$
We eliminate the Lagrange multipliers by differentiation, to get
$$
-g = \R_iT'(\ln x_i + 1) + \R_iT\rho_i'/\rho_i,~~ i = 1,2,\eqno(5.10)
$$
and  introduce the experimental evidence in the
form $\rho_1'/\rho_1 = \rho_2'/\rho_2$.

We begin the reduction of these equations by deriving 2 relations without
logarithms. The first is obtained by differentiation of Eq. (5.8),
$$
{\rho_1'\over\rho_1} = {\rho_2'\over \rho_2} = n {T'\over T},~~ n :=
{\sum \R_i \rho_i n_i \over \sum \R_i\rho_i}.
$$
The second relation is obtained by multiplying (5.10) by $\rho_i$, summing,
and combining with (5.8), and using the last result,
$$
-g(\rho_1 + \rho_2) = T\,'\sum\R_i\rho_i(n+1),
$$
or
$$
T\,' = {-g\mu \over \R(n+1)},~~ {1\over \mu}:=   {\sum \rho_i/\mu_i\over
\sum\rho_i} = 1/28.7179.
$$
Since the two densities are proportional, the quantities $n$ and $\mu$ are
uniform, and so is the temperature lapse rate.

We are now in position to evaluate the entropy.  Eq.s (5.10)  give us
$$
\ln x_i = (n+1)({\mu_i\over\mu}-1).~~ i = 1,2;
$$
also uniform. (values -.1026 and .3827, $x = .9024, y = 1.466$.)
Finally,
$$
k_{0i} = {\rho_i\over x_i} (\e T)^{-n_i}.
$$
with values $5.359\times10^{-11}$  and $ 1.011 \times
10^{-11}$, compare (5.5)  and $k_{02}/k_{01}=.1887$, see (5.1).

The entropies that are determined this way, using the observed constancy of
concentrations, are not very different from those calculated on the basis of the
Gibbs-Dalton hypothesis.

When heat is supplied or withdrawn, without loss or gain in total mass,
we note that
$$
k_{0i} = {p_i\over x_i\R_i} (\e T)^{-n_i-1},
$$
where $p_i$ is the partial pressure. Evaluated at ground level, the first
factor remains constant, so that the system moves on a path
$$
q := k_{01}^{n_2+1}/k_{02}^{n_1 + 1}  = {\rm constant}.
$$
The entropy density, given by the on shell condition (4.8) as
$$
s = \sum \R_i\rho_i\ln k_{0i} = \sum \R_i\rho_i(\ln \rho_i - n_i\ln \e T),
$$
is a function of the densities, the parameters and the temperature.
The temperature lapse rate (uniform) is 9.67 K per km and the
proportional rate at ground level $-T'/T(0)$ is $3.23\times 10^{-7}$ or
3.2 \% per km. The fractional density gradient is $\rho'(0)/\rho(0)) = 8.075 \%
$ per km.

\b \b

\ce{\bf V.5. Interaction model}

It is common to all models that failure of the Gibbs-Dalton hypothesis (about
equilibrium entropies) to account for the observation of (a single)
sound speed in mixtures is attributed to the effect of interactions between the
atoms of the two species. But the properties of mixtures at equilibrium do seem
to be accounted for by this hypothesis. It is therefore of interest to discover
whether the observed atmospheric concentrations can be accounted for without
affecting the entropies. A model of sound propagation examined in Section
IV.3 does have this property.

This first model has the appealing property of preserving the simplicity of the
Gibbs-Dalton equilibrium; the equation of motion that comes from variation of
$T$ has a unique solution such that the two terms in Eq.(4.4) vanish
individually; thus $x = y = 1$. The interaction term (4.11) affects the
density-variation equations. Instead of (5.2-3),
$$
\lambda_1 -\phi = \R_1 T(1 + n_1) +{\alpha\over 2}\sqrt{\rho_2\over \rho_1},
$$
$$
\lambda_2 -\phi = \R_2 T(1 + n_2) +{\alpha\over 2}\sqrt{\rho_1\over \rho_2}
$$
This is incompatible with constant concentration if $\mu_1 \neq\mu_2$.
This model cannot, by itself, do the job, but of course it may work in
conjunction with an ajustment of the entropy.

The second model leads to
$$
\lambda_1 -\phi = \R_1 T(1 + n_1) +{\alpha T\over 2}\sqrt{\rho_2\over \rho_1},
$$
$$
\lambda_2 -\phi = \R_2 T(1 + n_2) +{\alpha T\over 2}\sqrt{\rho_1\over \rho_2},
$$
and this is compatible with constant concentration for one value of  $\alpha$,
namely  $\beta = \alpha/4\R_1 \approx .004$.

\b\b
\ce{\bf 5.6. The hydrogen atmosphere}

Saha (1821) used his formula in  studies of dissociation in
stellar atmosphere, with  pressure and temperature data
obtained from other sources, with results that agreed well enough with
observation. But there was no study of these profiles within the
context of the theory. Here, with the advantage of a complete formulation of
the dynamics, we can calculate the profiles directly.

To describe a hydrogen  atmosphere we include the gravitational potential
in the lagrangian. An isolated atmosphere is characterized by
fixed entropy. The equilibrium relations (3.10-11) remain valid; they
are the two equations that remain when the
gravitational potential is eliminated   from the equations  of motion.
If instead we eliminate the logarithms we get the simple result given on
Eq.(5.11) below.

What is needed in addition is a
relationship between
$r$ and $T$ for constant entropy; that is, for constant values of
the parameters $k_{01}$ and $k_{02}$. The other equation of motion,
Eq. (3.8), is the fundamental relation of the gas, for which there is
no counterpart in Saha's approach. From it we can extract a formula for
the density. But the most revealing result is obtained by eliminating the
density
$\rho = \rho_1 +\rho_2$ from the three equations of motion. The result
is the following two simple equations of motion,
$$
\lambda - gz + \epsilon r = T<R(n+1)>,\eqno(5.11)
$$
and
$$
(n_1 - n_2) \ln T = {<\R(n+1)>\over \R} + (1 + r){\epsilon\over\R T} +
{\rm constant}.\eqno(5.12)
$$ where
$$
<R(n+1)> := \sum R_i(n_i + 1){\rho_i\over \rho}.
$$
Eq.(3.10) - resp. (3.11) - is a relation betweeen $r$ and $T$ that
holds through a sequence of equilibrium configurations with
different energies (assuming that $q$ is constant), at constant volume
- resp. at constant pressure.  Eq.(5.11-12), on the other hand, hold
throughout the atmosphere, with fixed entropy.

Eq. (5.11) is a smooth interpolation between the atmosphere of molecular
hydrogen at low temperatures and the atomic atmosphere of high
temperatures. There is an important change in the lapse
rate across the transition region.

The Lagrange multiplier $\lambda$ controls the total mass and the
constant in the second equation is the entropy; both are free parameters.

From (5.11) and (5.12) on easily obtains profiles of temperature and densities.
Further work in this direction is deferred.

\b\b
\ve
\no{\steptwo VI. Conclusions}

Let us assess what has been done and, especially, what has not been done
here.

(a)  It was, of course, well known that irrotational hydrodynamics can be
formulated as an eulerian field theory with an action principl.  Conservation
of energy is a cornerstone of thermodynamics and all adjacent sciences. What
appears to be new is including the temperature in the set of independent
dynamical variables, with respect to which the action is an extremum.
By this means all required information about a system is stored, once and for
all, in the lagrangian. Traditional thermodynamics is an on shell projection;
more precisely, a partial projection on the solutions of one or more  of the
variational equations.

(b) Given the gas law and the expression for the internal energy of an
ideal gas it is commonplace to calculate the entropy as a function of, say,
density and temperature. This is a fundamental relation of the ideal gas,
from which all its properties can be deduced (Callen 1960). What may be
new is that there is an off shell level of theory at which $S,T$ and
$\rho$ are independent variables, and that the fundamental relation is
the on shell condition obtained by variation of the temperature; it is
one of the equations of motion.

(c) The equations that constitute the Euler Lagrange equations do not
contain anything new, as far as the one component ideal gas is concerned.
What may be new is the total reliance on the chosen lagrangian for all
subsequent applications. Any expansion or generalization of the system or
of the context must be done (so we say) by making whatever changes that are
necessary in the lagrangian, in a manner so as not to upset any of the
applications that have already been successful. This last is a very strong
restraint on invention. It is the main reason for pursuing this line of
inquiry.  .

(d) Mixtures are usually studied by means of the Gibbs function, but such
studies are always, as far as we know, hampered by the fact that this
function is known only relatively to that of a reference configuration.
 It seems that
 the lagrangian gives more information than what has been
extracted from the Gibbs function. We have taken literally, perhaps more so
than is traditional, the additivity of lagrangians for composed systems.
We have been led to a  suggestion for  the lagrangians for mixtures that are
precise except for the values of a small set of parameters,
in number equal to the
number of components. This has greatly facilitated the task of using
experimental information to pin down the entropy of mixtures, something
that we have found to be unexpectedly easy. This is our best result.

(e) The suggestion  of ``second sound" in a gaseous mixture is not
confirmed by observation. We have suggested that this signals a limitation of
the Gibbs-Dalton hypothesis to serve as a guide for dynamics.
   It was shown that some of
the observations can be explained in therms of an interaction of densities,
without the assistance of kinetic theory. Further exploration of this model
must wait for more data, over a wider range of  frequencies and temperatures.

(f) There are seeds of a controversy in all this.  There is no
doubt that all the results can be obtained without using an action
principle.
 But, as was stressed already,  the action
principle is a powerful guide to further applications. Indeed, the action
principle, combined with what is known about gravitation, may be at variance
with the common conception of an isothermal equilibrium of an isolated gas in
the presence of a gravitational field. For this reason we have continued
 the investigation, begun in the first paper of this
series, of the classical arguments that seem to have
convinced all our thermodynamicists. Results have been relegated to the
Appendix, to avoid giving the impression that this is all we are trying to say
in this paper. We suggest   that
measurements of the temperature gradient in a supercentrifuge should be
carried out. Such measurements should
relatively simple; it would settle once and for all (via the principle of
equivalence) the question of the isolated atmosphere. A profound reassessment
would be required, whatever the result, for an isolated, isothermal atmosphere
does not have a natural place in thermodynamics (Fronsdal 2010), while the
existence of a temperature gradient is widely believed to be a violation of the
second law of thermodynamics.

\b\b\b
\no{\steptwo Acknowledgements}

I thank J. Rudnick and G Williams for very helpful conversations.

\ve

\no{\steptwo Appendix.  The controversial ``isolated" atmosphere}
\b
If a vertical column of an ideal gas is isolated from radiation, both
emitted and absorbed, then it may be reasonable to believe that both
may be left out of consideration. That is, an isolated column of gas in
the gravitational field would obey the equations of motion and exhibit a temperature gradient.
This conclusion was reached long ago by Loschmidt (1876) and others. It
was refuted by  Maxwell(1868) and Boltzmann(1909), as follows.

If we
would grant that an isolated, vertical column of air in equilibrium were
colder at the top and warmer at the bottom, then a paradox would arise.
Consider this arrangement: A heat bath with the lowest point at $z = 0$ is
maintained with temperature $T = T_0$ (at $z = 0$ if not everywhere).
Below it is suspended a vertical cylinder filled with an ideal gas, in
thermal contact at $z = 0$ with the heat bath, but otherwise isolated. The
theory predicts that, while the gas at the top of the cylinder has
temperature $T_0$, the bottom will be warmer, with temperature $T_1>T_0$.
It is proposed to utilize this temperature difference to run an engine,
taking heat from the bottom of the cylinder and returning it to the bath.
The argument is circular, for a complete analysis would require a theory
of heat engines operating between different levels of the gravitational
field. Maxwell circumvented this objection by replacing the heat engine by
a second cylinder, filled by another ideal gas, with different
characteristics (molecular weight, density, adiabatic index).
This second cylinder, at equilibrium, would have the same temperature
$T_0$ at the top, but a different temperature $T_2\neq T_1$ at the
bottom. Now provide thermal contact between the bottoms of the two
cylinders and suppose that $T_0<T_1<T_2$ at $t = 0$. Then both $T_1$ and
$T_2$ will begin to change, in the direction of equalization. If ``heat
flow" is defined in terms of the gradient of the temperature then,
according to Maxwell, a permanent, closed heat flow will be established;
apparently, a {\it perpetuum mobile} of the second kind.

We believe that the force of the conclusion is strongly affected if it
is shown that no energy flows.

\b

The {\it perpetuum
mobile} of Maxwell is one in which no work is being accomplished,
``nothing  really happens", since the energy flow of the final, asymptotic
configuration can be assumed to vanish without contradiction. But what
is this final state? Does it in fact exist? We believe that, at late
times, a stationary state will be reached, and we try to calculate it.

To remove a slight complication, let us take away the heat bath but retain
the thermal contact between the two cylinders at both ends, isolating the
system.  The original, adiabatic equations of motion cannot apply as they
stand, since they require two different lapse rates.  Another process is
needed, similar to ``cooling", in which there is a change in entropy. To
account for this additional process we use the model lagrangian  
$$
\L_{\rm tot} = \L[\Phi,\rho,T,...] + \L[k,\sigma],
$$
with the gravitational
potential included as before in the adiabatic part.  The field $\sigma$ is canonically conjugate to $T$. The domain is the
union of the two cylinders.    The equations of motion include the
equation of continuity, unchanged,  and,
at equilibrium,
$$
  -\phi + \lambda_i = \R_i T(\ln{k_i\over k_0} + 1),
~~ k_i := \rho_i/T^{n_i},~~ i = 1,2,
$$
$$
\R\rho_i(\ln{k_i\over k_0} - n_i) = c\,\Delta\sigma_i,
$$
$$
  c\,\Delta T = 0.
$$
It is important to emphasize that the entropy is not driven directly by
gravity - that would imply an unorthodox gravitational interaction and
probably a violation of the equivalence principle.  When the thermal
contact is made, between the lower ends of the two tubes, the
temperature at that point will eventually settle at a value $\hat T$
intermediary between $T_1$ and $T_2$, resulting in a temperature gradient
different from that induced by gravity in either of the two separated tubes.
Let
$$
\delta T'_i = (\hat T - T_i)/\ell,~~ i = 1,2,
$$
where $\ell$ is the length of the tubes, or more precisely the difference in
elevation between the upper and lower ends. This increment drives the
entropy, thus
$$
c\sigma_i''' = -{\R_i(n_i+1)\rho_i\over T}\delta T'_i.
$$
Because the process is reversible, $\sigma'''_1 + \sigma'''_2$ must be
zero, which leads to
$$
T
\,'\approx -g{\rho_1 +  \rho_2 \over    \R_1 \rho_1 (n_1 + 1)
+\R_2\rho_2 (n_2 + 1)},
$$
where, to this order, one should interpret the densities as averages.
This represents a reasonable
interpolation between the two extreme cases in which one or the other
cylinder is absent.

There is no energy flow within the tubes and, since the temperature is
continuous, no energy passes the boundary.

If instead one would justify the belief that a heat engine can be made to
run on the temperature difference between the lower ends of the two
cylinders (Graeff 2007), then a similar analysis involving entropy needs to
be made.
\b\b\b

\no{\steptwo References}

\no Bardeen, J.M., A variational principle for rotating stars
in General Relativity,

 Astrophys. J. {162}, 7 (1970).

\no Bernoulli, D.,   Argentorat, 1738.

\no Callen, H.B.,
{\it Thermodynamics; an introduction to the physical theories of
equilibrium

thermostatics and irreversible thermodynamics},
 Wiley, NY 1960.

\no DeHoff, P., {\it Thermodynamics in Material Science}, McGraw-Hill NY
1993.

\no de la Mora, J.F. and Puri,  A., Two fluid Euler theory,  of sound dispersion in gas mixtures
of disparate masses, J. Fluid Mec. {\bf 168} 369-382 (1985).

\no Eddington, A.S., {\it The internal constitution of stars}, Dover, N.Y.
1959.

\no Eggert, J., Uber den Dissozitionzustand der Fixstern-gase,

Physik.
Zeitschr. XX, 570-4, 1919.

\no Emden,  {\it Gaskugeln}, Teubner 1907.

 Physik.
Zeitschr. XX, 570-4, 1919.

\no Emden,  {\it Gaskugeln}, Teubner 1907.

\no Fetter, A.L. and Walecka, J.D., {\it Theoretical Mechanics of Particles
and Continua},

McGraw-Hill, N.Y. 1980.

\no  Fourier, J.B.J., {\it Th\'eorie analytique de la chaleur}, Didot, Paris
1822.

\no Fronsdal, C., Reissner-Nordstrom and charged polytropes,

Lett.Math.Phys. {\bf 82}, 255-273 (2007).

\no Fronsdal, C., Stability of polytropes, Phys.Rev.D, 104019 (2008).

\no Fronsdal, C., ``Heat and Gravitation. II. Stability", arXiv 0904.0427.

\no Fronsdal, C., Heat and gravitation I. The Action principle,\ submitted for
publication, 

arXiv

\no Gay-Lussac, J.L.,  ~The Expansion of Gases by Heat,  Annales de chimie
43 , 137- (1802).

\no Gibbs, W., On the Equilibrium of Heterogeneous Substances,
 Trans. Conn. Acad.,  1875.

 \no Guldberg, CM and Waage,  P.,   Studies Concerning Affinity, Forhandlinger:

 Videnskabs-Selskabet i Christinia {\bf 35} , 1864

\no Graeff, R.W., Viewing the controversy Loschmidt-Boltzman/Maxwell
through macroscopic

measurements of the temperature gradient in vertical
columns of water, preprint, 2008.

\no Holman, J.P., {\it Thermodynamics}, McGraw-Hill NY 1974.
 
\no Holmes, R. and Tempest, W., The Propagation of Sound in Monatomic Gas
Mixtures,   

898-904 (1960).
 
\no Lane, H.J., On the Theoretical Temperature of the Sun, under the
Hypothesis of a gaseous

Mass maintaining the Volume by its interbal Heat,
and depending on the laws of gases

 as known to terrestrial Experiment,
Amer.J.Sci.Arts, Series 2, {\bf 4}, 57- (1870).

\no Laplace, P.S.,   Paris 1816.

\no Loschmidt, L., Sitzungsb. Math.-Naturw.
Klasse Kais. Akad. Wissen. {\bf 73.2} 135 (1876).

 \no Maxwell, J.C., The London, Edinburgh and Dublin Philosophical Magazine
{\bf35} 215 (1868).

\no M\"uller, I., {\it A History of Thermodynamics}, Springer, Berlin 2007.

\no Poisson, S.D., {\it Th\'eorie mathªmatique de la chaleur},  1835.

\no Putterman, S.J., ~~Superfluid Hydrodynamics", North HJolland 1974.

\ve
\no Ritter, A., A series of papers in Wiedemann Annalen, now Annalen der
Physik,

 For a list see Chandrasekhar (1938). The volumes 5-20 in
Wiedemann

Annalen appear as the volumes 241-256 in Annalen der Physik.

 \no Saha, M.N., Ionization in the solar atmosphere, Phil. Mag. {\bf
40}, 472-488 (1920).

   \no Schutz, B.F., Perfect fluids in General Relativity; velocity
potentials and a variational

principle, Phys. Rev.  {\bf D2} 2762-2771
(1770)
 
\no Taub, A.H., Generally relativistic variational principle for perfect
fluids,

Phys. Rev. {\bf 94}, 1468 (1954).
 
\no Tolman, R.C., {\it Relativity, Thermodynamics and Cosmology},
Clarendon, Oxford 1934.
\ve

\end